\documentclass[aps,prx,reprint,preprintnumbers,superscriptaddress,nofootinbib,longbibliography,floatfix]{revtex4-2}

\usepackage[utf8]{inputenc}
\usepackage[T1]{fontenc}
\usepackage{lmodern}
\usepackage{microtype}
\usepackage{mathtools}

\usepackage{placeins}

\usepackage{amsmath}

\newcommand{\nocontentsline}[3]{}
\newcommand{\tocless}[2]{\bgroup\let\addcontentsline=\nocontentsline#1{#2}\egroup}

\usepackage[caption=false]{subfig}
\usepackage{amsfonts}
\usepackage{graphicx}
\usepackage{hyperref}
\hypersetup{
  colorlinks=true,
  citecolor=blue,
  linkcolor=blue,
  urlcolor=blue
}

\usepackage{color}
\definecolor{darkred}{rgb}{1.0,0.1,0.1}
\definecolor{darkgreen}{rgb}{0.1,0.7,0.1}
\definecolor{darkblue}{rgb}{0.1,0.1,1.0}

\DeclareRobustCommand{\Sec}[1]{Sec.~\ref{sec:#1}}

\DeclareRobustCommand{\App}[1]{App.~\ref{app:#1}}

\DeclareRobustCommand{\Fig}[1]{Fig.~\ref{fig:#1}}
\DeclareRobustCommand{\Figs}[2]{Figs.~\ref{fig:#1} and \ref{fig:#2}}
\DeclareRobustCommand{\Eq}[1]{Eq.~(\ref{eq:#1})}
\DeclareRobustCommand{\Eqs}[2]{Eqs.~(\ref{eq:#1}) and (\ref{eq:#2})}
\DeclareRobustCommand{\Eqss}[3]{Eqs.~(\ref{eq:#1}), (\ref{eq:#2}), and (\ref{eq:#3})}

\DeclareRobustCommand{\Ref}[1]{Ref.~\cite{#1}}

\DeclareRobustCommand{\Refs}[1]{Refs.~\cite{#1}}

\begin{document}

\preprint{MIT-CTP 5315}

\title{Neural Conditional Reweighting}

\author{Benjamin Nachman}
\email{bpnachman@lbl.gov}
\affiliation{Physics Division, Lawrence Berkeley National Laboratory, Berkeley, CA 94720, USA}
\affiliation{Berkeley Institute for Data Science, University of California, Berkeley, CA 94720, USA}

\author{Jesse Thaler}
\email{jthaler@mit.edu}
\affiliation{Center for Theoretical Physics, Massachusetts Institute of Technology, Cambridge, MA 02139, USA}
\affiliation{The NSF AI Institute for Artificial Intelligence and Fundamental Interactions}

\begin{abstract}
There is a growing use of neural network classifiers as unbinned, high-dimensional (and variable-dimensional) reweighting functions.
To date, the focus has been on marginal reweighting, where a subset of features are used for reweighting while all other features are integrated over.
There are some situations, though, where it is preferable to condition on auxiliary features instead of marginalizing over them.
In this paper, we introduce neural conditional reweighting, which extends neural marginal reweighting to the conditional case.
This approach is particularly relevant in high-energy physics experiments for reweighting detector effects conditioned on particle-level truth information.
We leverage a custom loss function that not only allows us to achieve neural conditional reweighting through a single training procedure, but also yields sensible interpolation even in the presence of phase space holes.
As a specific example, we apply neural conditional reweighting to the energy response of high-energy jets, which could be used to improve the modeling of physics objects in parametrized fast simulation packages.
\end{abstract}

\maketitle


{\small
\tableofcontents
}

\section{Introduction}

A common task in particle physics is to reweight one set of events $P$ to match the statistical properties of another set of events $Q$.
Here, $P=\{x_i\}$, $x_i\in\mathbb{R}^N$, are drawn independently and identically distributed from probability density $p(x)$, and $Q$ is similarly drawn from $q(x)$.
The reweighting function,
\begin{equation}
    w(x)\approx \frac{q(x)}{p(x)}\,,
\end{equation}
ensures that the expectation value of any weighted observable computed from $P$ will match the same value computed from $Q$ on average.%
\footnote{In certain cases, it is possible to resample the events to match the statistical uncertainties as well~\cite{Nachman:2020fff}.}
For example, $P$ could be events from a control region while $Q$ are events from a signal region, or $P$ could be from simulation while $Q$ could be from data, or $P$ and $Q$ could be from two different simulations with different parameter choices.

In nearly every case of interest in particle physics, $p$ and $q$ are not known analytically. 
When $x$ is low-dimensional, it is common to create histograms to estimate $p$ and $q$ from the events in $P$ and $Q$.
One can then construct a binned reweighting function by taking ratios of the bin contents.
This works well when $w(x)$ is slowly varying and $x$ is low- (and fixed-) dimensional.
When these conditions are not met, the traditional binned approach is not effective.

Neural network classifiers can be used to form unbinned, high- (and variable-) dimensional reweighting functions, which can be viewed as an application of simulation-based inference (see \Ref{2019arXiv191101429C} for a review).
In particular, the optimal classifier for distinguishing events drawn from $P$ and $Q$ is (any monotonic function of) the likelihood ratio $q(x)/p(x)$.
Therefore, one can approximate $w(x)$ directly by interpreting the output of a classifier trained to distinguish the two event samples.
This feature of classifiers is well known~\cite{hastie01statisticallearning,sugiyama_suzuki_kanamori_2012} and has been widely used in particle physics for parameter estimation~\cite{Cranmer:2015bka,Brehmer:2018kdj,Brehmer:2018eca,Brehmer:2019xox,Brehmer:2018hga,Stoye:2018ovl,1907.08209,Hollingsworth:2020kjg,2010.03569}, domain adaptation~\cite{Andreassen:2020nkr}, detector parametrizations~\cite{Badiali:2020wal}, and unfolding~\cite{bunse2018unification,Ruhe2019MiningFS,Andreassen:2019cjw,Andreassen:2021zzk}.

\begin{figure*}
    \centering
    \includegraphics[width=0.81\textwidth]{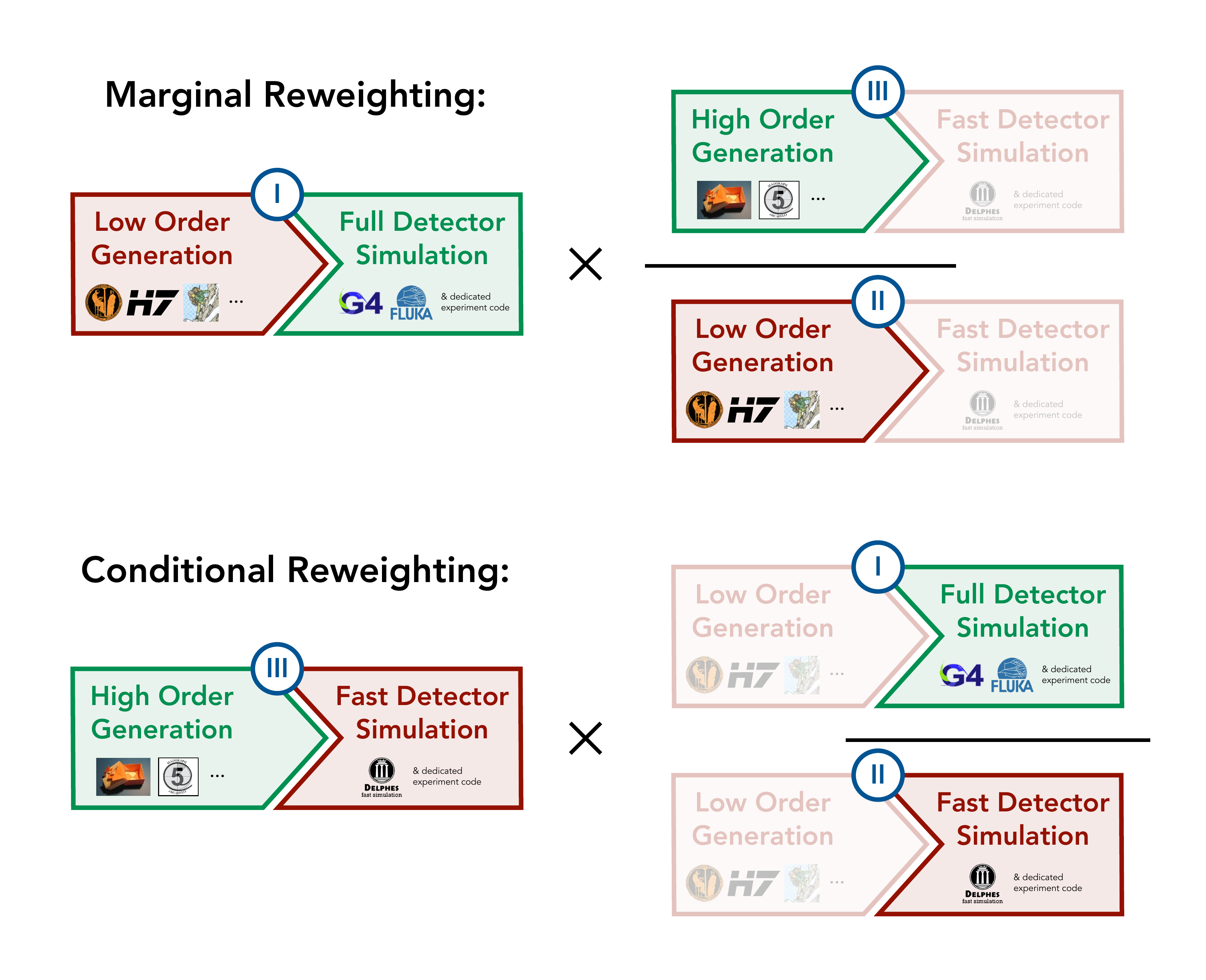}
    \caption{
    Schematic diagrams contrasting marginal reweighting (top) with conditional reweighting (bottom), in the context of generation and simulation for collider physics.
    The goal is to create an event sample that has the particle-level kinematics of a high-order generator (e.g.\ \textsc{Powheg-Box}~\cite{Alioli:2010xd,Nason:2004rx,Frixione:2007vw} or \textsc{MG5\_aMC}~\cite{Alwall:2014hca}) with the detector-level reconstruction of a full detector simulation (based on e.g.\ \textsc{Geant}~4~\cite{Agostinelli:2002hh,1610988,Allison:2016lfl} or \textsc{Fluka}~\cite{Battistoni:2015epi,BOHLEN2014211}).
    In marginal reweighting, one reweights events from a low-order generator (e.g.\ \textsc{Pythia}~\cite{Sjostrand:2006za,Sjostrand:2014zea}, \textsc{Herwig}~\cite{Bahr:2008pv,Bellm:2015jjp}, or \textsc{Sherpa}~\cite{Gleisberg:2008ta,Sherpa:2019gpd}) to match the kinematics of a high-order generator, marginalizing over the simulator.
    In conditional reweighting, one reweights events from a fast simulation (e.g.\ based on \textsc{Delphes}~\cite{deFavereau:2013fsa,Mertens:2015kba,Selvaggi:2014mya}) to match the reconstruction of a full detector simulation, conditioning on the generator.
    }
    \label{fig:schematic}
\end{figure*}

To our knowledge, in all applications to date of classifier-based reweighting, other event features $x'$ are integrated over, such that:
\begin{equation}
    w(x)\approx \frac{\int dx'\, q(x,x')}{\int dx' \, p(x,x')}\,.
\end{equation}
This marginalization is often necessary when $x'$ is not observable, as is the case when $x$ represents detector-level quantities and $x'$ represents particle-level quantities.
This can be an issue, however, if $w(x)$ is applied to another data set where the probability density of $x'$ is not the same as $q(x')$.
For example, suppose that $x'$ represents the particle-level jet energy, $x$ is the detector-level jet energy, $p(x)$ represents the probability density of a fast simulation, and $q(x)$ is the probability density for a full simulation.
One can train a model to reweight $p(x)$ to $q(x)$ to match the detector resolution, but if $p(x')\neq q(x')$, then there is a degeneracy between physics and detector effects.
Even if $p(x')=q(x')$ (or if one reweights $x$ and $x'$ simultaneously), the reweighting function cannot be applied to another data set with a different energy distribution.
It would therefore be ideal to reweight the conditional probabilities instead, such that:
\begin{equation}
\label{eq:cond_reweight}
    w(x)\approx \frac{q(x|x')}{p(x|x')}.
\end{equation}

In this paper, we introduce neural conditional reweighting, which is a strategy to extract the conditional probability ratio in \Eq{cond_reweight}.
We first show how to achieve conditional reweighting by training two independent classifiers, one for joint reweighting and one for marginal reweighting. 
We then develop a custom loss function specifically for conditional reweighting, which is better suited to situations with phase space holes.
Through a single training procedure, the resulting neural network can sensibly interpolate across minimally populated regions of phase space.
We demonstrate the efficacy of our approach using simple Gaussian examples and a more realistic application in collider physics.

The primary motivating application of neural conditional reweighting is shown in \Fig{schematic}, where the goal is to improve generation and simulation for collider physics.%
\footnote{To avoid overlap in word usage, we use the word ``generator'' to refer to particle-level simulation tools, and ``simulator'' to refer to detector-level simulation tools.}
Here, we have three synthetic data sets:
\begin{itemize}
    \item (I):  Coarse Generator $\Rightarrow$ Precise Simulator;
    \item (II):  Coarse Generator $\Rightarrow$ Coarse Simulator;
    \item (III):  Precise Generator $\Rightarrow$ Coarse Simulator.
\end{itemize}
Data sets (I) and (II) use a coarse particle-level generator while data set (III) uses a precise particle-level generator.
By contrast, data sets (II) and (III) use a coarse detector-level simulator while data set (I) uses a precise detector-level simulator.
The goal is to create a data set that has the most precise particle-level generation and the most precise detector-level simulation, which requires merging the best features of data sets (III) and (I), respectively.
One way to construct this merged data set is to perform a marginal reweighting from the coarse particle-level truth to the precise particle-level truth, shown in the top line of \Fig{schematic}.  
Here, we advocate for conditional reweighting, where we reweight only the detector response from (II) to (I) and then apply this to data set (III), shown in the bottom line of \Fig{schematic}.

In the limit of infinite statistics and no phase space holes, both marginal reweighting and conditional reweighting yield the same final distributions.
The aim of this paper is to highlight situations where conditional reweighting could outperform marginal reweighting in practical situations.
In principle, one could bypass data set (II) entirely and directly conditional reweight (III) to (I), but we will argue that this is likely never better than marginal reweighting.
Beyond reweighting, once can train surrogate models for generation and simulation (see \Refs{Butter:2020tvl,Alanazi:2021grv} for reviews), which we do not consider here.

\begin{figure}[t]
    \centering
    \includegraphics[width=0.5\textwidth]{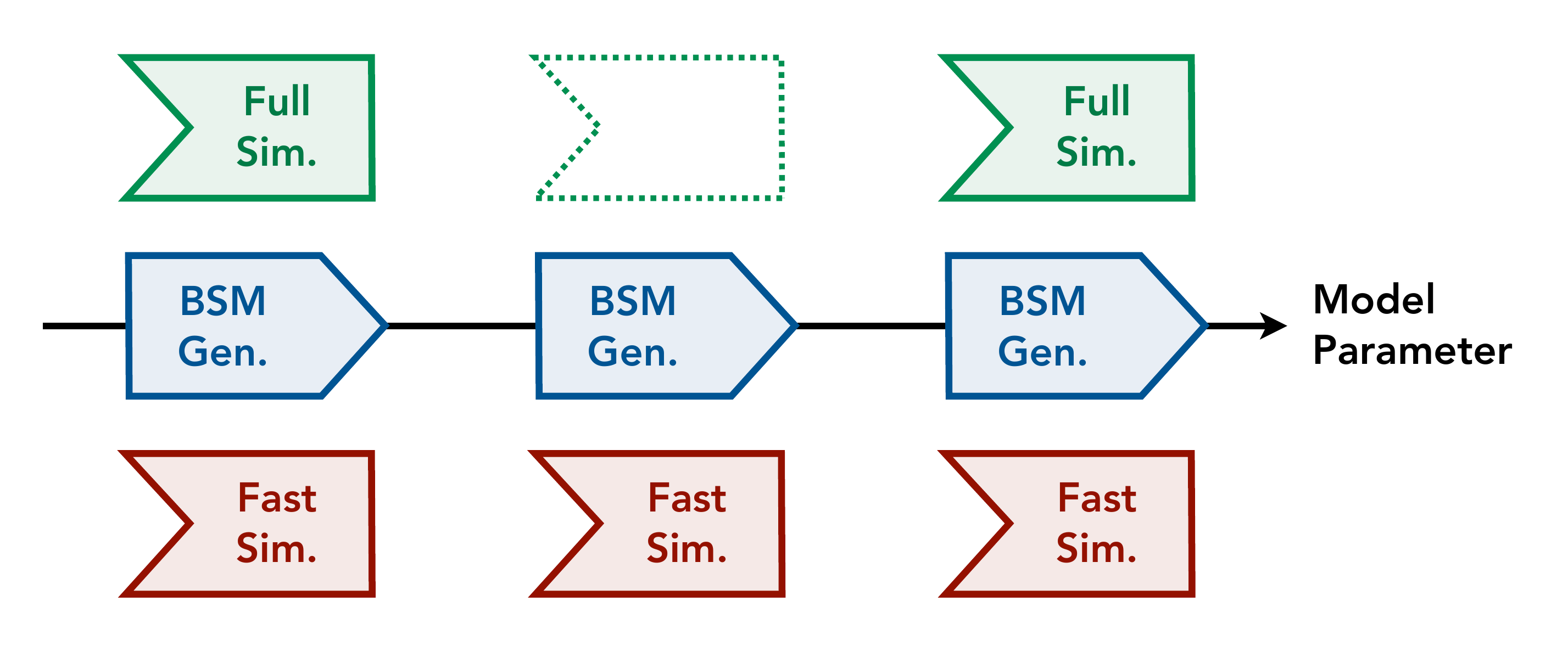}
    \caption{An illustration of how neural conditional reweighting could be used for search for physics beyond the Standard Model (BSM).
    Here, BSM samples are fully simulated (Full Sim.)\ only for a small number of model parameter values (usually particles masses).  One can emulate fully simulated samples (dotted polygon in the top row) by using conditional reweighting from BSM generator samples (BSM Gen.)\ combined with fast simulation (Fast Sim.).}
    \label{fig:BSM}
\end{figure}

While we have framed our discussion in terms of the motivating example above, there are many other potential use cases for conditional reweighting in high-energy physics.
Another illustrative example is for new physics searches, as shown in \Fig{BSM}.
In this case, full simulation data sets may only be available at benchmark signal parameter values.
By using conditional reweighting, one can interpolate between these signal benchmarks with the help of fast simulation.
While it is possible to interpolate limits at the level of model parameters, interpolation across regions of rapidly changing kinematic properties due to phase space constraints can be difficult, and
conditional reweighting could be more accurate in this context.
Another potential example is scale factors to correct simulation to agree better with data.
Typically, reweighting is not conditional on auxiliary features that may differ between the calibration sample and test sample.
There are cases where scale factors derived with neural networks are parametrized~\cite{Andreassen:2020nkr}, but these existing methods also reweight the conditional variable, whereas this would be avoided with conditional reweighting.

The remainder of this paper is organized as follows.
In \Sec{math}, we review neural reweighting and generalize the marginal version to the conditional case.
We present a simple Gaussian example to illustrate the complementary of conditional and marginal reweighting in \Sec{gaussian}.
In \Sec{jer}, we present an application of neural conditional reweighting in the context of jet energy measurements at the Large Hadron Collider (LHC).
The paper ends with our conclusions and outlook in \Sec{conclusions}.


\section{The Statistics of Conditional Reweighting}
\label{sec:math}

\subsection{Review of Marginal Reweighting}

Let $f:\mathbb{R}^N \rightarrow [0,1]$ be a classifier with the goal of distinguishing events generated by probability densities $p$ and $q$.
This function can be obtained by minimizing an appropriate loss functional, such as the binary cross entropy (BCE):
\begin{align}
\label{eq:bce}
L_{\rm BCE}[f] =-\int dx\, \Big( & p(x) \log f(x) + q(x)  \log (1-f(x)) \Big)\,.
\end{align}
In practice, with finite training data, we would replace
\begin{align}
    \int dx \,p(x)\Rightarrow \sum_{x_i \in P}\,,
\end{align}
but for the remainder of this discussion, we consider the infinite statistics limit such that we can replace sums over events by integrals and then use functional optimization to determine the optimal classifier $f$.

The function $f_\text{BCE}$ that optimizes the functional in \Eq{bce} has the following well-known property~\cite{hastie01statisticallearning,sugiyama_suzuki_kanamori_2012}: 
\begin{align}
\label{eq:f_to_p}
    \frac{1-f_\text{BCE}(x)}{f_\text{BCE}(x)}=\frac{q(x)}{p(x)}\,,
\end{align}
such that one learns the per-instance likelihood ratio in the asymptotic limit.
Note that this analysis assumes the same number of events sampled from $q$ and $p$; if these are not the same, then \Eq{f_to_p} is multiplied by the relative frequency of the two random variables (prior ratio).
Similar formulae apply to other loss functionals, and certain loss functionals such as the maximum likelihood classifier (MLC) loss~\cite{DAgnolo:2018cun,2101.07263} result in classifiers that directly approximate the likelihood ratio without the transformation in \Eq{f_to_p}.

In the case that the feature space consists of observed features $x \in \mathbb{R}^N$ and hidden (or latent) features $x' \in \mathbb{R}^M$, but the classifier $f$ is only a function of $x$, then the learned function is related to the marginalized likelihood ratio:
\begin{equation}
    \label{eq:marginal_reweighting}
    \frac{1-f_\text{BCE}(x)}{f_\text{BCE}(x)} = \frac{q(x)}{p(x)} \equiv \frac{\int dx' \, q(x,x')}{\int dx' \, p(x,x')}.
\end{equation}
We call this procedure \emph{marginal reweighting}.
Note that the same symbols $p$ and $q$ are used to denote the marginal (e.g.~$p(x),p(x')$) and joint (e.g.~$p(x,x')$) probability densities, and primes are used to separate observed and latent quantities.

If, instead, we consider a classifier $f:\mathbb{R}^{N+M} \rightarrow [0,1]$ that depends on the full $(N+M)$-dimensional feature space, then the optimal learned function is related to the joint likelihood ratio:
\begin{equation}
    \label{eq:joint_reweighting}
    \frac{1-f_\text{BCE}(x,x')}{f_\text{BCE}(x,x')} = \frac{q(x,x')}{p(x,x')},
\end{equation}
and we call this \emph{joint reweighting}.

A challenge faced by marginal (and to a lesser extent joint) reweighting is that the weights can become large and unphysical if $q(x)$ and $p(x)$ do not have overlapping support.
In particular, if there is a region of phase space where $p(x) \simeq 0$, then \Eq{marginal_reweighting} becomes singular.
When this happens, conditional reweighting offers an alternative reweighting strategy.

\subsection{Conditional Reweighting with Two Classifiers}
\label{sec:cond_two}

We can easily extend the above formalism to conditional reweighting by noting the following:
\begin{align}
\label{eq:conditional_reweighting}
    \frac{q(x|x')}{p(x|x')} \equiv \frac{\frac{q(x,x')}{q(x')}}{\frac{p(x,x')}{p(x')}}=\frac{q(x,x')}{p(x,x')}\,\frac{p(x')}{q(x')}\,.
\end{align}
The first term is the joint reweighting in \Eq{joint_reweighting}.
The second term is the inverse of the marginal reweighting in \Eq{marginal_reweighting}, with the roles of $x$ and $x'$ reversed.
Therefore, one can achieve conditional reweighting with two functions, each trained as a standard classifier.

A potential challenge with applying \Eq{conditional_reweighting} in practice is that $q(x')$ might have inadequate support relative to $p(x')$ in some regions of phase space, leading to ill-behaved weights.
Note that this is effectively the opposite problem as faced by marginal reweighting, so it is typically less of an issue in practice.
That said, we can partially mitigate this issue by leveraging the ability of neural networks to interpolate.

\subsection{Conditional Reweighting with a Single Classifier}
\label{sec:ncr_single}

A natural question is whether conditional reweighting could be learned in one learning step, instead of in two steps as in the above construction.
A somewhat trivial way to accomplish this is to note that
\begin{align}
\label{eq:yprimelimit1}
    \frac{q(x|x')}{p(x|x')} = \lim_{y' \to x'} \frac{q(x,x')\, p(y')}{p(x,x') \, q(y')},
\end{align}
where $y' \in \mathbb{R}^M$.
Therefore, to learn this ratio, we could train a classifier $f(x,x',y')$ to distinguish \emph{pairs} of events drawn from $p(x,x') \, q(y')$ versus $q(x,x')\, p(y')$, and then set $x' = y'$.
The reason this is somewhat trivial is that, assuming the BCE loss, the optimal classifier factorizes into two separate classifiers,
\begin{align}
\label{eq:double_bce}
    \frac{1-f_\text{BCE}(x,x',y')}{f_\text{BCE}(x,x',y')}
    =
    \frac{1-g_\text{BCE}(x,x')}{g_\text{BCE}(x,x')}
    \, 
    \frac{h_\text{BCE}(y')}{1-h_\text{BCE}(y')}\,,
\end{align}
which might as well be optimized separately as in \Eq{conditional_reweighting}.

A more interesting construction follows from the relation:
\begin{align}
\label{eq:yprimelimit2}
    \frac{q(x|x')}{p(x|x')} = \lim_{y' \to x'} \frac{q(x,y')\, p(x')}{p(x,x') \, q(y')}\,,
\end{align}
where the primed arguments in the numerator have been flipped relative to \Eq{yprimelimit1}.
Before taking $x' = y'$, we can learn this ratio with a new neural conditional reweighting (NCR) loss functional: 
\begin{align}
\label{eq:ncr}
L_{\rm NCR}[f] = & -\int dx \,dx'\,dy\,dy'\, p(x,x') \, q(y,y')\\
& ~ \times \Big( \log f(x,x',y') + \log (1-f(y,x',y')) \Big)\,.
\nonumber   \end{align}
Swapping the $x$ and $y$ integral labels in the second term, it is straightforward to show that the optimal classifier is:
\begin{align}
\label{eq:ncr_solution}
\frac{1-f_\text{NCR}(x,x',y')}{f_\text{NCR}(x,x',y')} = \frac{q(x,y') \int dy \, p(y,x')}{p(x,x') \int dy \, q(y,y')}\,.
\end{align}
Inserting this into \Eq{yprimelimit2}, we find
\begin{align}
\label{eq:ncr_final}
\frac{1-f_\text{NCR}(x,x',x')}{f_\text{NCR}(x,x',x')} =      \frac{q(x|x')}{p(x|x')}\,,
\end{align}
which is our default approach to conditional reweighting.%
\footnote{The approach in \Eq{yprimelimit1} corresponds to replacing $f(y,x',y')$ with $f(y,y',x')$ in the second term of \Eq{ncr}.}

The NCR loss in \Eq{ncr} is similar to the BCE loss in \Eq{bce}, but instead of the events sampled from $p$ and $q$ contributing separately to the two terms, the events contribute to both terms.\footnote{This form is similar to the setup in \Ref{pmlr-v80-belghazi18a,2101.07263,Kim:2021pcz} where events are combined to use deep learning as way to estimate mutual information. See \Ref{Miller:2021hys} for a related construction.}
For $x,y\in\mathbb{R}^N$ and $x',y'\in\mathbb{R}^M$, the density $p(x,x')\,q(y,y')$ means that a $(2N+2M)$-dimensional data set is sampled from $p$ and $q$ independently.
In practice, one can approximate \Eq{ncr} by using the standard BCE loss with one $(N+2M)$-dimensional data set sampled from $p(x,x')\,q(y')$ with a label of $1$ and a second data set sampled from $p(x')\,q(y,y')$ with a label of $0$.

Because \Eq{ncr_final} is obtained from the $y' \to x'$ limit, we can partially mitigate the issue from \Sec{cond_two} of ill-behaved weights when there are dead regions of phase space.
This works because neural networks typically yield sensible and smooth interpolations across the training domain, so we can use $y' \not= x'$ information to predict the behavior at $y'=x'$.
As shown in \App{alt}, we find only modest differences between using one classifier trained using the NCR loss (\Eq{ncr_final}) and two classifiers each trained with the BCE loss (\Eq{double_bce}).
We find larger differences when comparing conditional reweighting against marginal reweighting, where the issue of dead phase space regions is more pronounced.

In the example application shown in \Fig{schematic}, conditional reweighting is learned from two data sets (I) and (II) that have the same low-order generator, which means that $p(x') = q(x')$.
From the above derivation, though, we see that this restriction is not necessary, and conditional reweighting can be learned for any $p(x')$ and $q(x')$ with overlapping support, a fact we leverage in \Sec{jer}.
In practice, though, it is helpful for $p(x')$ and $q(x')$ to be similar, not only to ensure overlapping support but also to avoid unnecessarily large weights
This is the reason why we recommend using data set (II) to derive the conditional reweighting factor in \Fig{schematic}, instead of directly conditional reweighting (III) to (I).

\subsection{Technical Implementation}

In our subsequent case studies, the functions $f$ trained with the NCR loss will be parametrized with neural networks.
While it is possible to train a generic function $f(x,x',y')$, we can take advantage of the known form of the optimal solution.
Rewriting \Eq{ncr_solution}, the optimal $f$ takes the form
\begin{equation}
    f_{\rm NCR}(x,x',y') = \frac{p(x|x')}{p(x|x')+q(x|y')}.
\end{equation}
Interestingly, although $f$ is naively a function of three variables, the optimal function can be expressed in terms of two functions of two variables each.

Armed with this insight, we construct our classifiers as
\begin{align}
\label{eq:condtwofuncs}
    f(x,x',y') = \frac{e^{f_0(x,x')}}{e^{f_0(x,x')}+e^{f_1(x,y')}}\,,
\end{align}
where $f_0,f_1$ are each neural networks.
The exponential is used because each term must be non-negative (as a conditional probability density).
In fact, since $f_0$ and $f_1$ are expected to be similar log likelihoods, we can further simplify the problem by building $f_0$ and $f_1$ from the same components:
\begin{align}\label{eq:f0}
    f_0(x,x')&=W_0'\max(0,W_0 \, g(x,x') + b_0)+b_0',\\\label{eq:f1}
    f_1(x,y')&=W_1'\max(0,W_1 \, g(x,y') + b_1)+b_1'\,,
\end{align}
where $g:\mathbb{R}^{N+M}\rightarrow\mathbb{R}^{L_0}$ is a neural network, $W_i\in\mathbb{R}^{L_0\times L_1}, W_i'\in\mathbb{R}^{L_1\times 1}$ are weight matrices, and $b_i\in\mathbb{R}^{L_1},b_i'\in\mathbb{R}$ are biases.
In other words, $f_0$ and $f_1$ are shallow neural networks with a single hidden layer of size $L_1$ with the Rectified Linear Unit (ReLU) activation function that takes as input a common deep neural network that outputs size $L_0$.

All neural networks are implemented using \textsc{Keras}~\cite{keras} with the \textsc{Tensorflow} backend~\cite{tensorflow} and optimized with \textsc{Adam}~\cite{adam}.
Because we use BCE-like loss functions, \Eq{f_to_p} is needed to convert the classifier output to a likelihood ratio.
The marginal reweighting networks consist of three hidden layers with 50 nodes per layer.
The ReLU activation function is used for the intermediate layers while a sigmoid activation is used for the last layer.
Each network is trained for 50 epochs with early stopping using a patience of 10 and deploys a batch size of 1000.
Conditional reweighting uses the same training schedule and a similar network architecture: the $g$ function in \Eqs{f0}{f1} has two hidden layers with 50 nodes per layer and the ReLU activation.
The shallow $f_i$ networks have $L_1=50$.

\section{Gaussian Examples}
\label{sec:gaussian}

We now present simple numerical examples that explore when conditional reweighting may be as good as or superior to marginal reweighting.
Here, each data set in \Fig{schematic} is a one-dimensional Gaussian random variable.
The ``particle-level truth'' random variables $T_i$ are described by means $\mu_i$ and standard deviations $\sigma_i$ with:
\begin{align}
    \mu_0&\equiv\mu_{\text{(I)}}=\mu_{\text{(II)}},&
    \sigma_0 &\equiv\sigma_{\text{(I)}}=\sigma_{\text{(II)}},\\
    \mu_1&\equiv\mu_{\text{(III)}},&
    \sigma_1&\equiv\sigma_{\text{(III)}}.
\end{align}
The corresponding ``detector-level reconstructed'' random variables $R_i$ are given by
\begin{align}
    R_i=T_i+Z_i,
\end{align}
where $Z_i$ is a Gaussian random variable with mean $b_i$ and standard deviation $\epsilon_i,$ with:
\begin{align}
    b_0 &\equiv b_{\text{(I)}},&
    \epsilon_0&\equiv\epsilon_{\text{(I)}},\\
    b_1 &\equiv b_{\text{(II)}}=b_{\text{(III)}},&
    \epsilon_1& \equiv\epsilon_{\text{(II)}}=\epsilon_{\text{(III)}}.
\end{align}
The desired target distribution combines the generation parameters of data set (III) with the simulation parameters of (I):
\begin{align}
  \mu_{\rm target} &= \mu_1, & \sigma_{\rm target} &= \sigma_1,\\
  \quad b_{\rm target} &= b_0, & \epsilon_{\rm target} &= \epsilon_0.
\end{align}

In each of the examples below, one million events are used for each data set with a 50\% test-train split.
In the collider physics context, one could use more events from data sets (II) and (III), as they are computationally cheaper to produce than (I), which involves the full simulation.
None of these parameters were optimized, but we find that the results are stable to small changes in the setup.
For conditional reweighting, the reweighter in \Eqss{condtwofuncs}{f0}{f1} is trained using data sets (I) and (II) and then applied to data set (III); alternative implementations of conditional reweighting are shown in \App{alt}.

\subsection{Overlapping Support}

\begin{figure*}[p]
    \centering
    \includegraphics[width=0.4\textwidth]{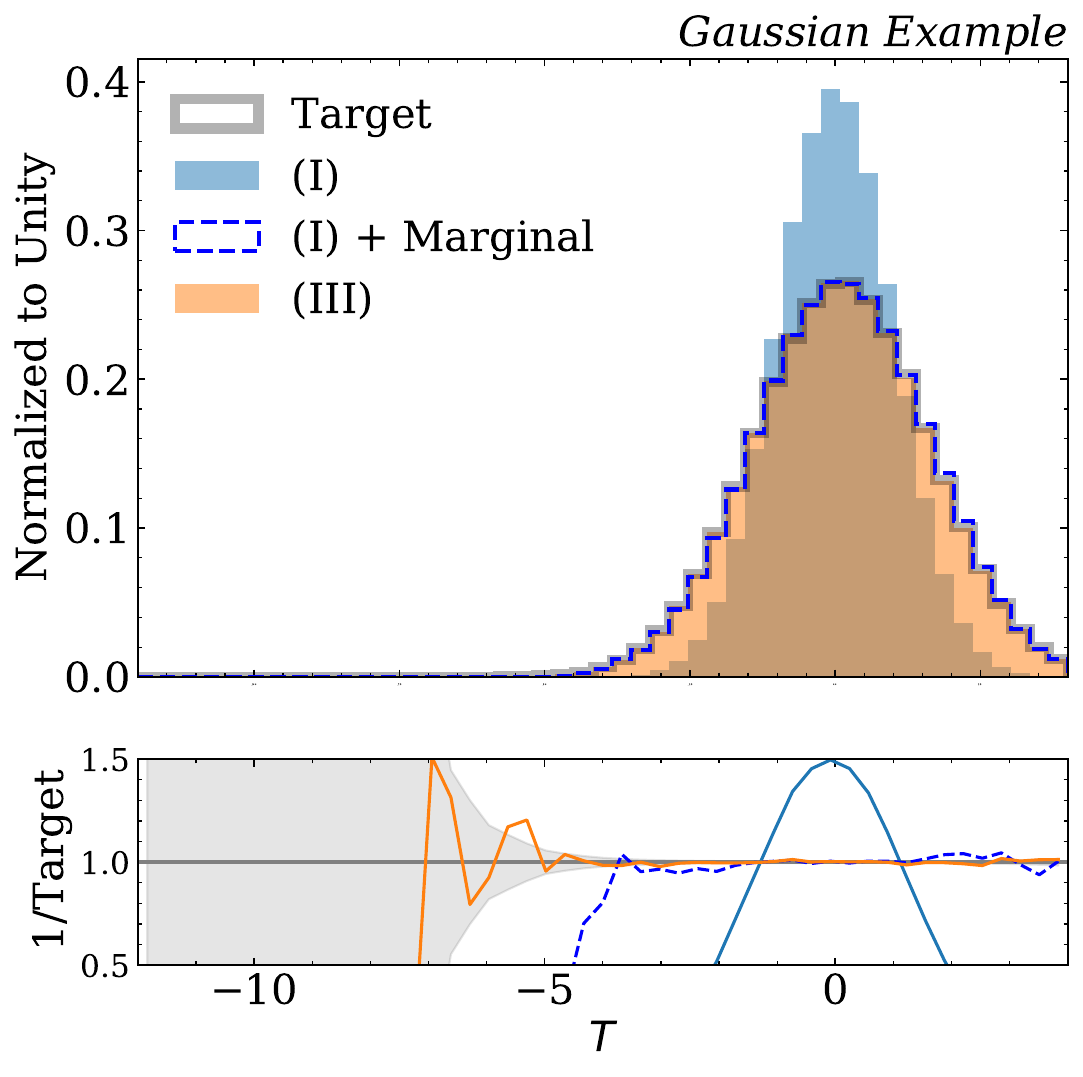} $\qquad$
    \includegraphics[width=0.4\textwidth]{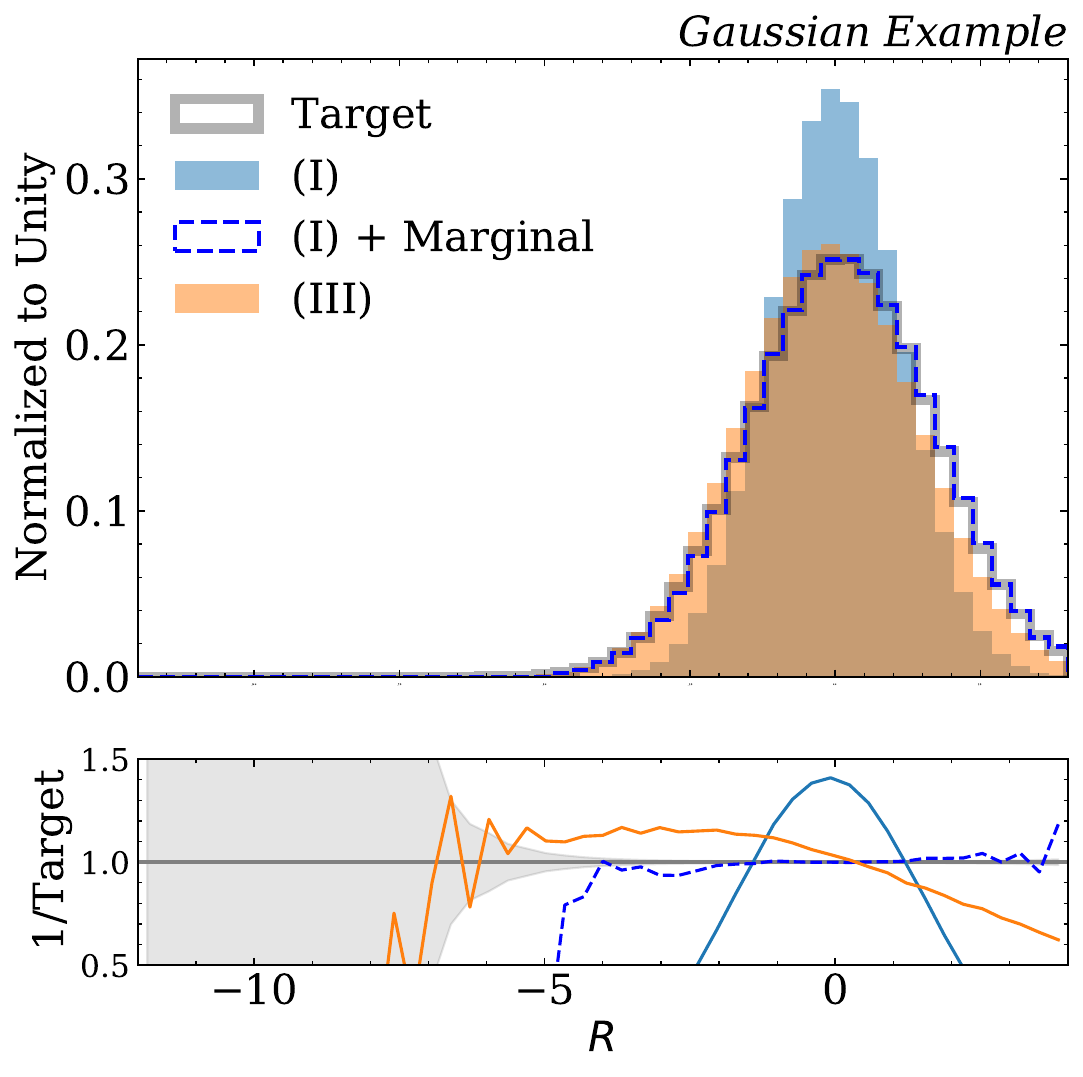}
    
    \includegraphics[width=0.4\textwidth]{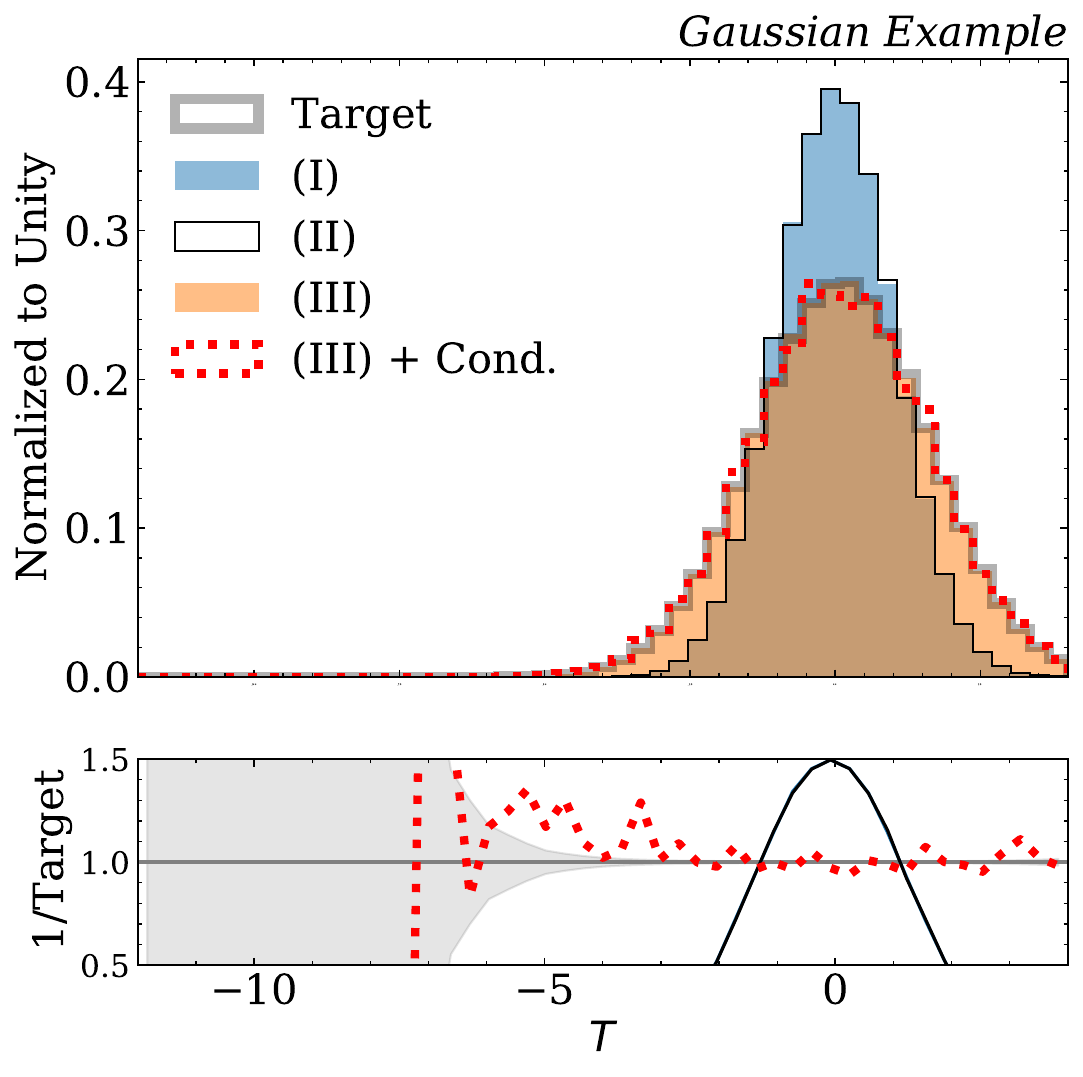} $\qquad$
    \includegraphics[width=0.4\textwidth]{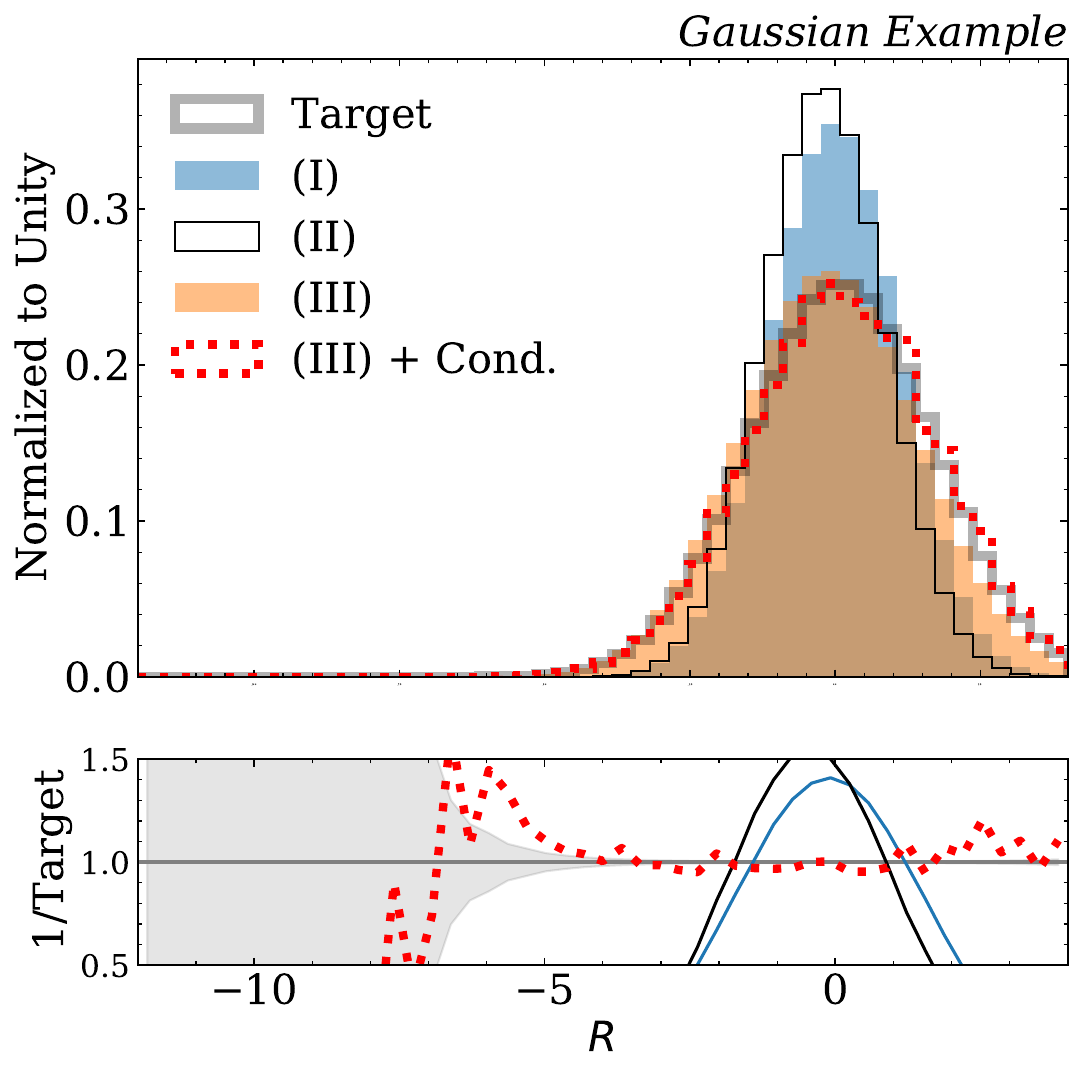}
    \caption{
    Comparison of marginal reweighting (top row) and conditional reweighting (bottom row) in a plain Gaussian example.
    Histograms of the random variables are shown at ``particle level'' (left column) and ``detector level'' (right  column).
    Distribution (I) involves a coarse generator interfaced with a precise simulator.
    Distribution (II) involves a coarse generator interfaced with a coarse simulator.
    Distribution (III) involves a precise generator interfaced with a coarse simulator.
    To match the target distribution (precise generator interfaced with a precise simulator), one can either marginally reweight distribution (I) or conditionally reweight distribution (III).
    In this case, marginal reweighting yields better performance than conditional reweighting.
    }
    \label{fig:gaussian}
\end{figure*}

\begin{figure*}[t]
    \centering
    \includegraphics[width=0.4\textwidth]{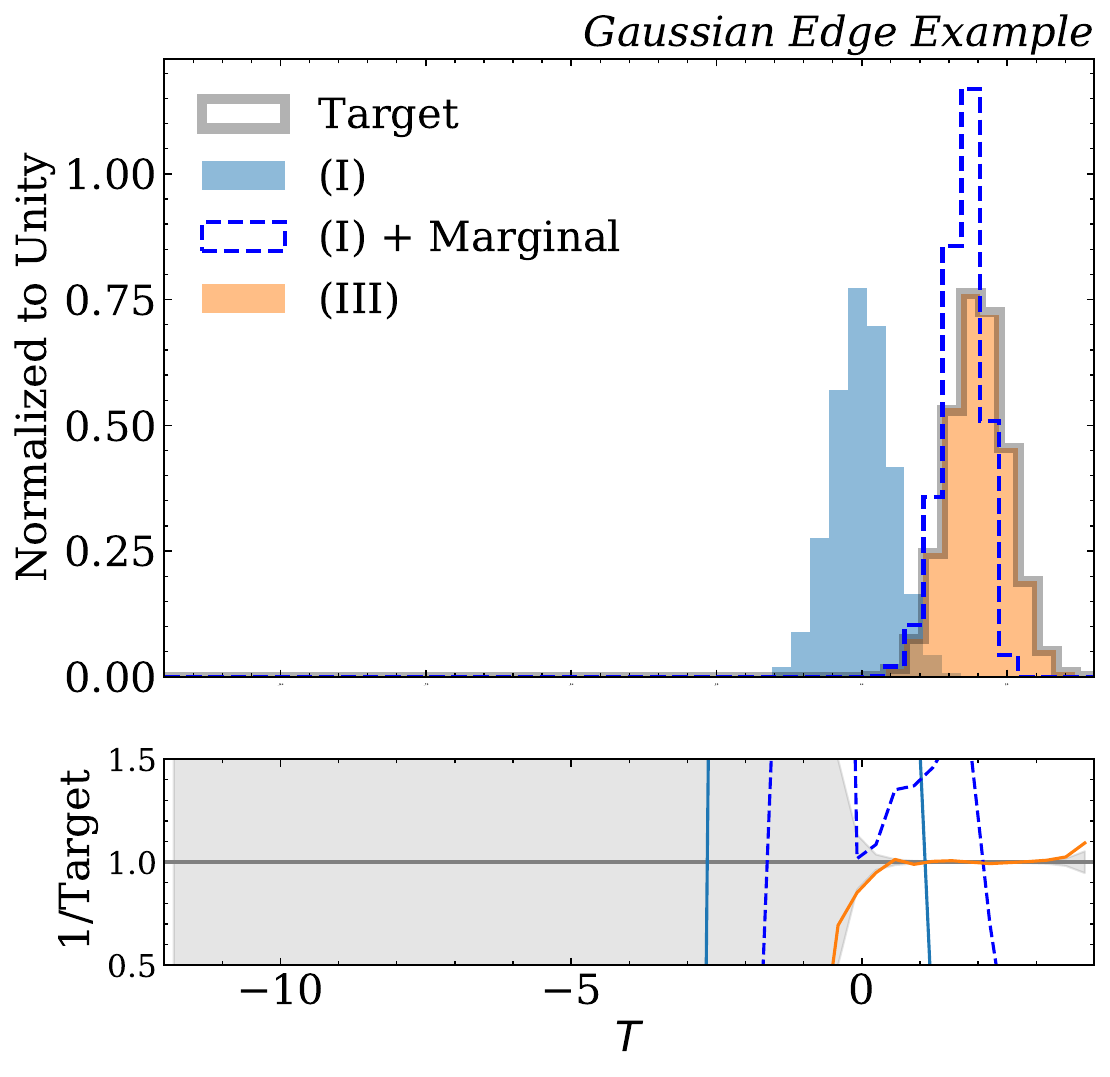} $\qquad$
    \includegraphics[width=0.4\textwidth]{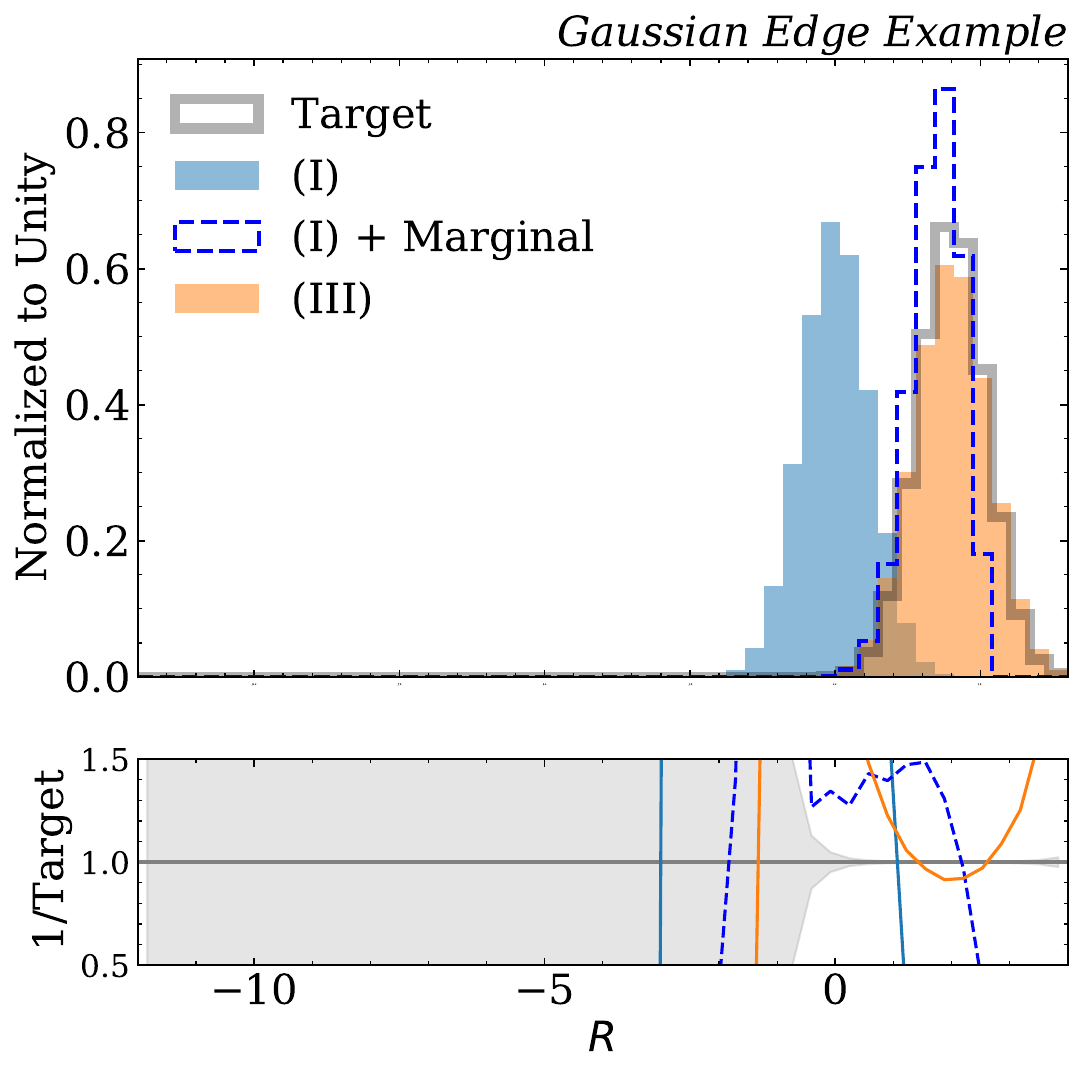}
    
    \includegraphics[width=0.4\textwidth]{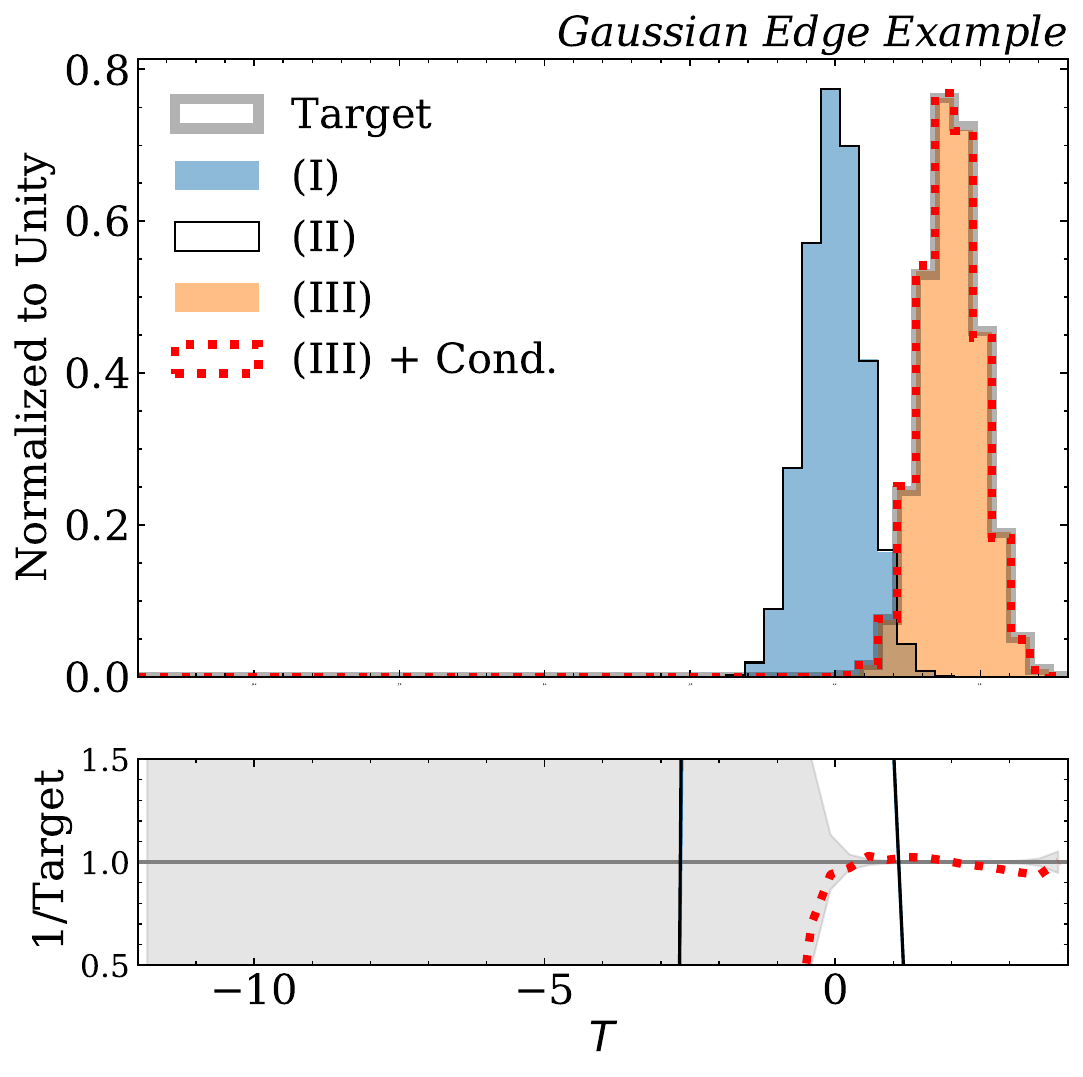}$\qquad$
    \includegraphics[width=0.4\textwidth]{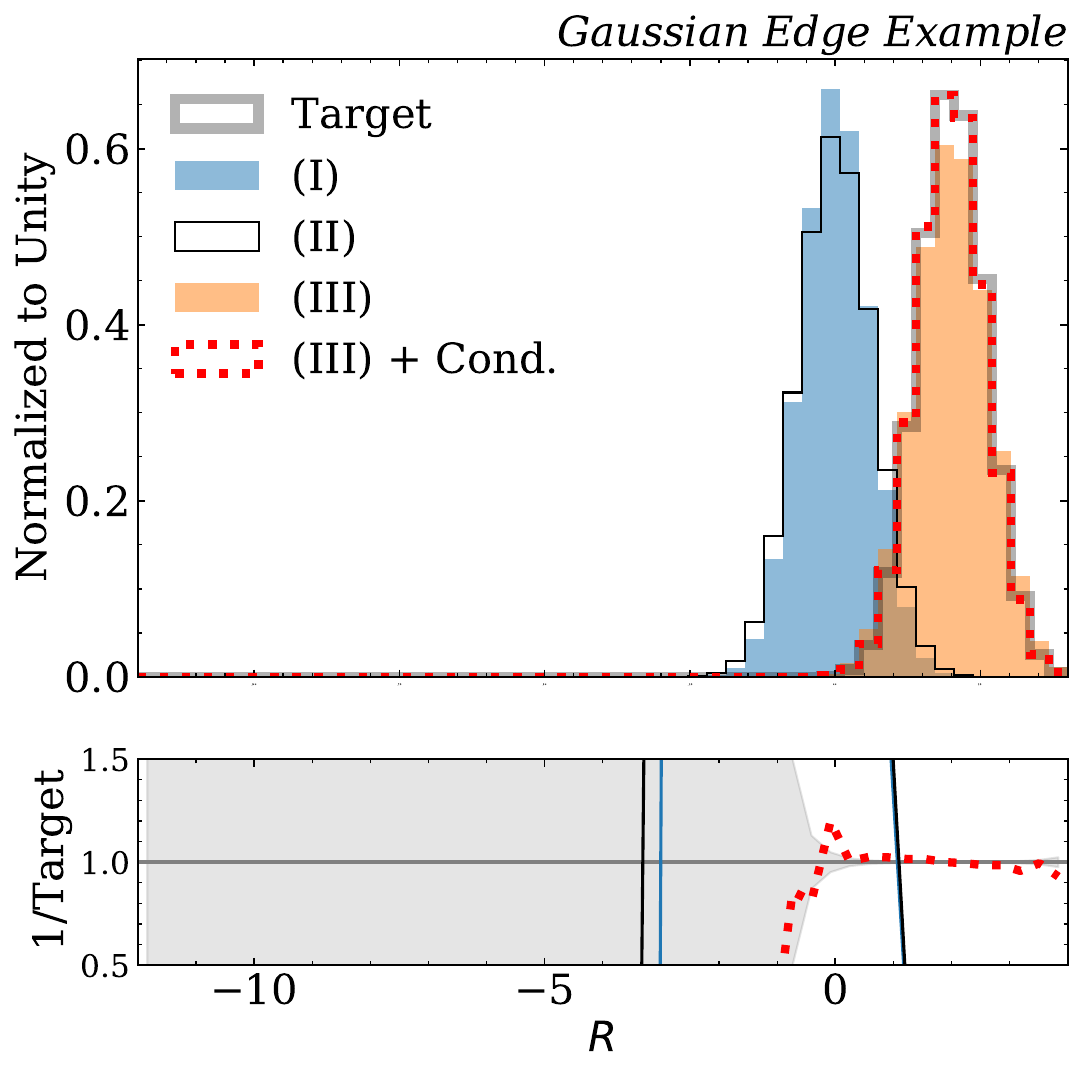}  
    \caption{
    Same as \Fig{gaussian}, but for a Gaussian edge example involving
    extrapolation.
    Marginal reweighting only yields sensible results where distributions (I) and (III) have significant phase space overlap.
    Conditional reweighting, by contrast, is able to extrapolate outside the naive training domain.
    }
    \label{fig:gaussianEx}
\end{figure*}

\begin{figure*}[t]
    \centering
    \includegraphics[width=0.4\textwidth]{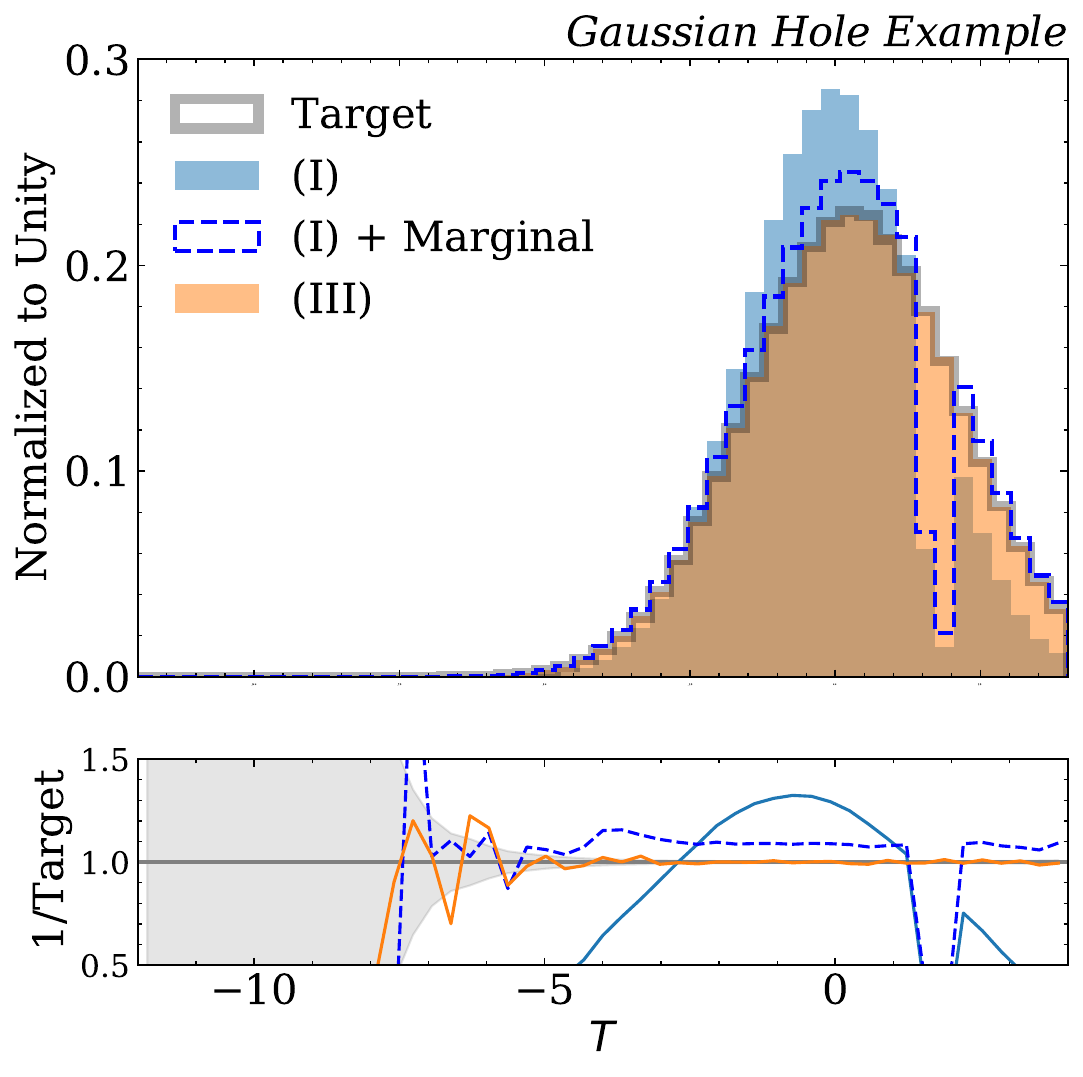}$\qquad$
    \includegraphics[width=0.4\textwidth]{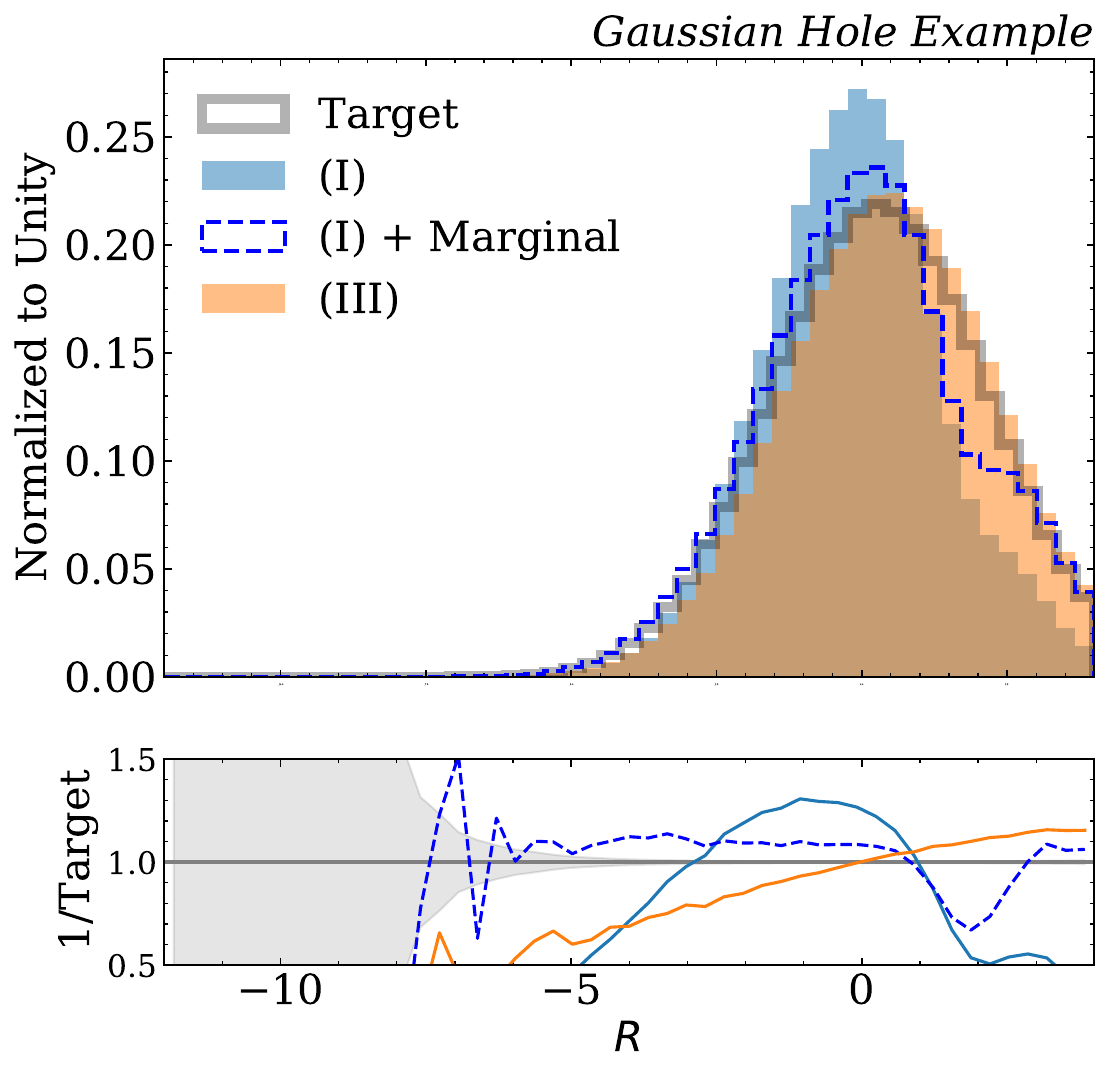}
    
    \includegraphics[width=0.4\textwidth]{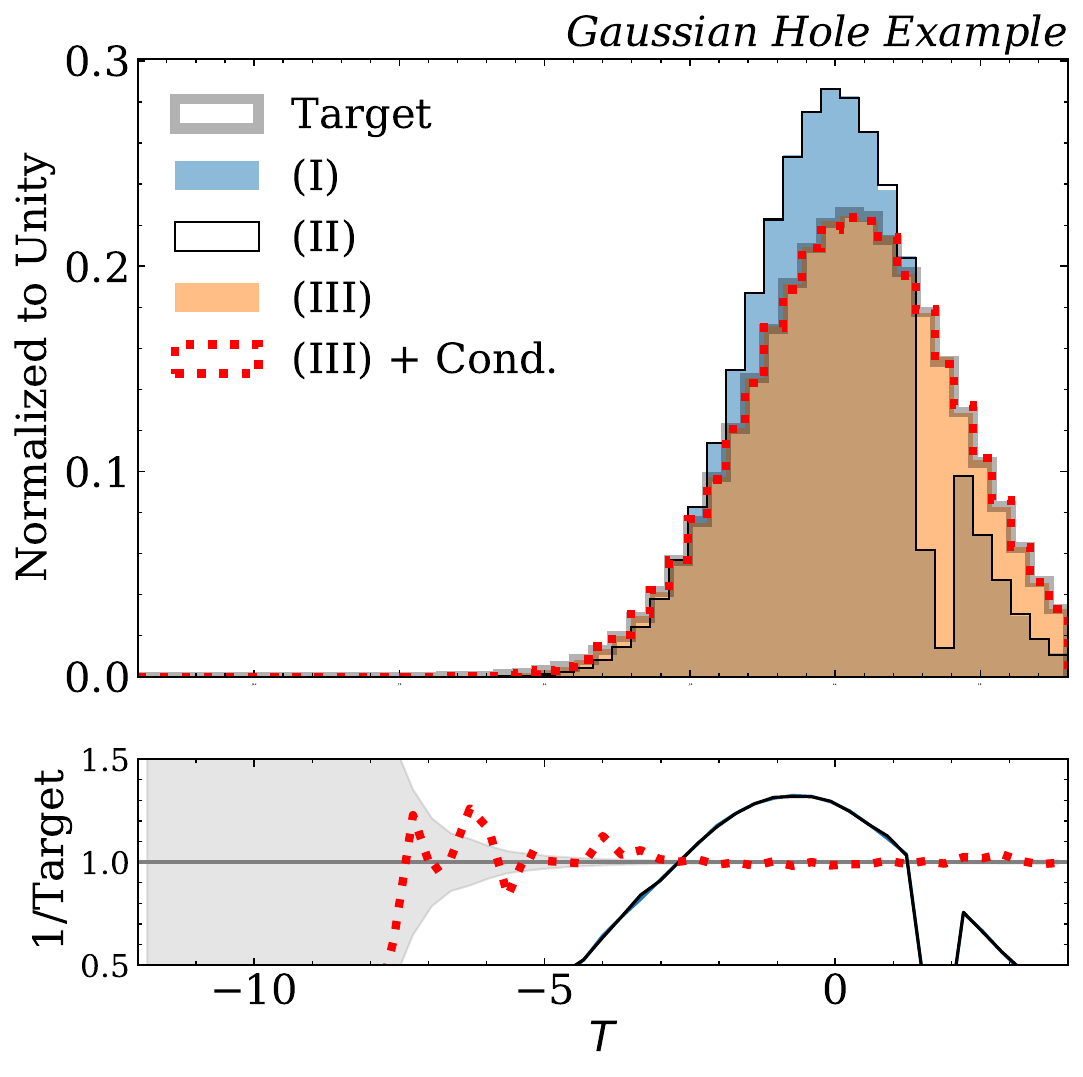}$\qquad$
    \includegraphics[width=0.4\textwidth]{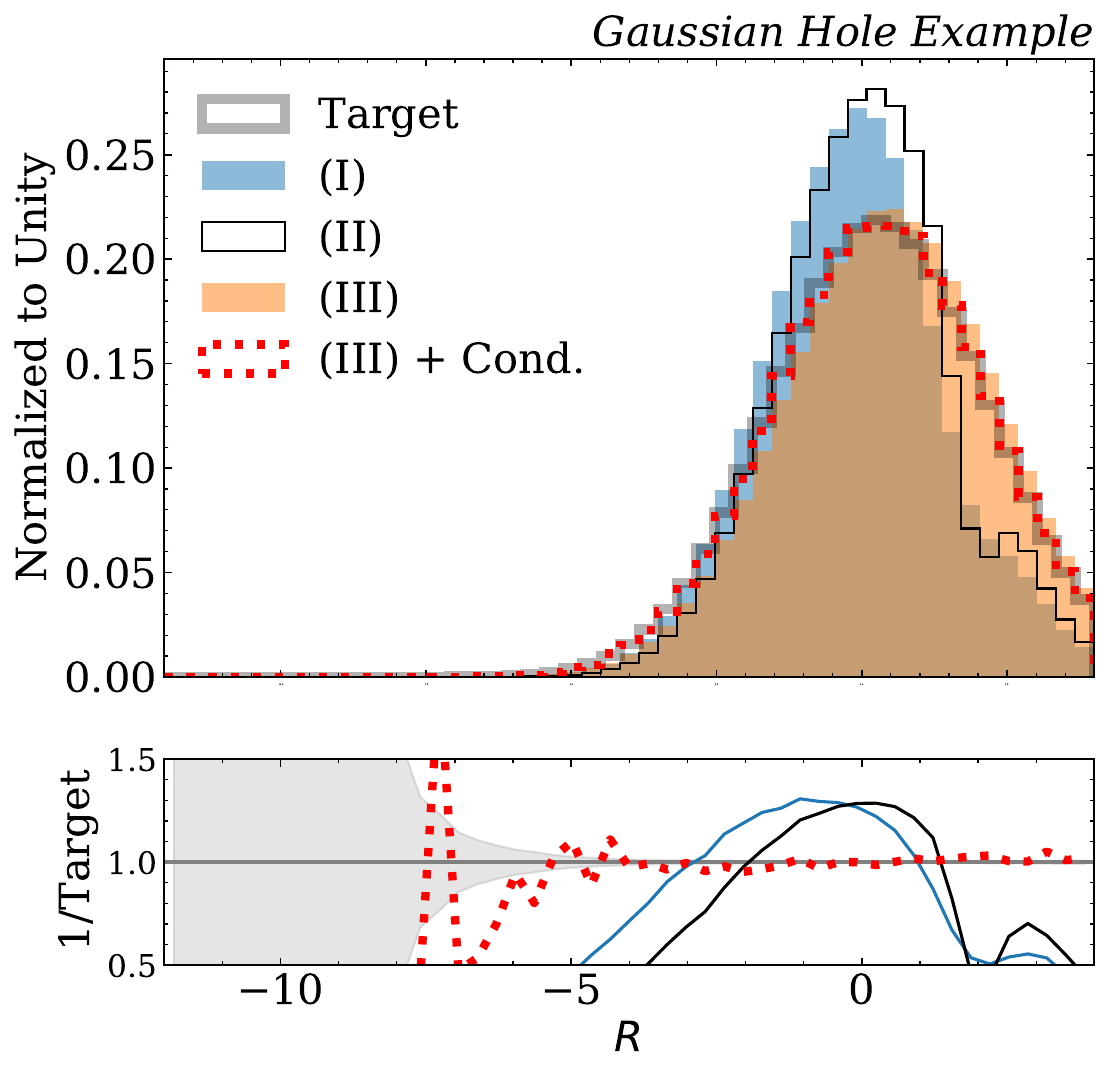}
    \caption{
    Same as \Fig{gaussian}, but for a Gaussian hole example involving
    interpolation.
    Whereas marginal reweighting cannot accurately model the phase space gap, conditional reweighting is able to sensibly interpolate.
    }
    \label{fig:gaussianIn}
\end{figure*}

For our first example, we consider a situation with no phase space gaps, such that marginal reweighting is expected to already perform well.
The particle-level generation parameters are
\begin{equation}
    \mu_0=0.0, \quad \mu_1=0.1, \quad \sigma_0=1.0, \quad \sigma_1=1.5,
\end{equation}
such that the distributions have overlapping support.
The detector-level simulation parameters are
\begin{equation}
    b_0=0.0, \quad b_1=-0.2, \quad \epsilon_0=0.5, \quad \epsilon_1=0.3,
\end{equation}
so that the distortions are relatively small.

The results of marginal and conditional reweighting are shown in the top row and bottom row of \Fig{gaussian}, respectively.
The left column shows the truth distribution for $T_i$ while the right column shows the reconstructed distribution for $R_i$.
Both marginal and conditional reweighting are able to achieve the target distribution.
In particular, the conditionally reweighted (III) and the marginally reweighted (I) have the same particle-level distributions as data set (III), but the detector response of data set (I).

Upon careful inspection, one can see that marginal reweighting is able to match the target distribution more precisely than conditional reweighting, especially in the tails of the Gaussians.
The exact agreement with the target varies with different pseudoexperiments and with different random initializations of the networks. 
However, this trend is robust:  marginal reweighting is more precise than conditional reweighting in this context.
This is to be expected because the conditional path in \Fig{schematic} requires a more complicated setup and involves a higher-dimensional learning problem compared to marginal reweighting.
Without any phase space gaps, the marginal reweighting strategy in \Eq{marginal_reweighting} is sufficient.

\subsection{Extrapolation}

To understand a context where conditional reweighting might be able to outperform marginal reweighting, consider the situation where there is a large hierarchy in the particle-level truth distributions:
\begin{equation}
    p_{\text{(I)}}(T) \ll p_{\text{(III)}}(T).
\end{equation}
In this case, the marginal weights in \Eq{marginal_reweighting} can become very large.
Conditional reweighting may also fail in this context, but if the truth distribution in data set (II) is chosen to be close to the truth in data set (I), then at least one does not encounter large weights.
To the extent that neural networks can sensibly extrapolate outside of the training domain, one can then conditionally reweight the reconstructed data set (III) to the target distribution.
Of course, one has to do a careful validation in any situation that involves extrapolation.

To explore the performance of conditional reweighting for extrapolation, the particle-level generation parameters for this example are:
\begin{equation}
\mu_0=0.0, \quad \mu_1=2.0, \quad \sigma_0 = \sigma_1 = 0.5,
\end{equation}
such that there is very little overlap in their support, with the mean of the (III) truth being four standard deviations away from the mean of the (I) truth.
On the other hand, the detector-level simulation parameters are:
\begin{equation}
b_0=b_1=0.0, \quad \epsilon_0=0.3, \quad \epsilon_1=0.4,
\end{equation}
such that the difference in the smearing behavior is relatively small.

The performance of marginal and conditional reweighting for extrapolation is presented in \Fig{gaussianEx}, with the same layout as \Fig{gaussian}.
For $T\lesssim 1$, where $p_{\text{(I)}}(T)\gtrsim p_{\text{(III)}}(T)$, marginal reweighting is effective for the particle-level truth.
For $T\gtrsim 1$, though, marginal reweighting fails to reproduce the target distribution because there are either few or no events in data set (I) to upweight.
These same trends are present at detector-level in the upper right plot of \Fig{gaussianEx}.

By contrast, conditional reweighting is able to match both the truth and reconstructed distributions.
Because detector effects in this case are so similar between data sets (I) and (II), the conditional reweighting is nearly constant.
For the particle-level truth, the good agreement is more or less guaranteed by construction, since data set (III) already has the desired truth distribution.
For the detector-level reconstruction, there is no information to constrain the conditional reweighting for $R\gtrsim 1$, but the neural network is nevertheless able to smoothly extrapolate from the region of phase space with plenty of events.

We therefore expect that when the reweighting function is constant or smoothly continues from its behavior in regions with high density overlap, conditional reweighting may be as precise as it is in this example.
This example also motivates explicitly prioritizing smooth extrapolation as part of the training loss.

\subsection{Interpolation}
\label{sec:gauss_interpolation}

Neural networks are known to be effective at interpolating, which is in general less fraught than extrapolation.
There are known cases where regions of phase space may be undercovered by certain generators (e.g.\ dead zones in \textsc{Herwig} 7~\cite{Reichelt:2017hts}) or where a reweighting derived from one process needs to be applied to another with significantly different phase space distributions.
This will be the context that we study for the jet energy response example in \Sec{jer}.

To study the interpolation case for the Gaussian example, we start with generation parameters
\begin{equation}
\label{eq:truth_para_interpolation}
\mu_0=0.0, \quad \mu_1=0.3, \quad \sigma_0=1.5, \quad \sigma_1=1.8,
\end{equation}
and simulation parameters
\begin{equation}
    b_0=0.0,\quad b_1=0.2, \quad \epsilon_0=0.5, \quad \epsilon_1=0.3,
\end{equation}
such that there is good phase space overlap.
Then, we introduce a modification of the model, where
\begin{equation}
\Pr(|T_0-c| < \delta)=0 \quad  \text{for} \quad c=1.75, \quad \delta=0.25.
\end{equation}
Apart from this modification, the probability density of $T_0$ is proportional to a Gaussian distribution with the stated parameters in \Eq{truth_para_interpolation}.
This creates a gap in phase space that necessitates interpolation.

The performance of marginal and conditional reweighting for interpolation is shown in \Fig{gaussianIn}, again with the same layout as \Fig{gaussian}.
Similarly to extrapolation, marginal reweighting at truth level is very effective away from the gap in phase space.
Since $p_{\text{(I)}}(T)=0$ in the gap, however, it is impossible for marginal reweighting to match the target distribution, for which the probability density is non-zero.
This carries over to detector level, where marginal reweighting suffers near $R\sim 1$.
By contrast, conditional reweighting is effective across the entire domain, albeit with worse precision than marginal reweighting far from the phase space gap.

\section{Jet Energy Response}
\label{sec:jer}

We now present a physics case study where we expect conditional reweighting to be effective: simulation of the jet energy response at the LHC.
To highlight the performance of conditional reweighting for interpolation, we will artificially construct a large phase space gap.
Since we do not have a full target distribution to compare with the reweighted distributions, we use marginal reweighting on a sample without the phase space gap as a proxy.
Despite these limitations, we hope this example highlights the complementarity of marginal and conditional reweighting.

\subsection{Simulated Dijet Data Sets}

This study is based on generic dijet production in quantum chromodynamics.
As the ``coarse'' particle-level generator for data set (I), we use \textsc{Pythia}~6.426~\cite{Sjostrand:2006za} with the Z2 tune~\cite{Chatrchyan:2011id}.
The ``precise'' particle-level generator for data set (III) is \textsc{Pythia}~8.219~\cite{Sjostrand:2007gs}.
Different from the study in \Sec{gaussian}, we also use the ``precise'' \textsc{Pythia}~8.219 for data set (II), though as described below, we impose a phase space restriction such that data sets (I) and (II) have similar phase space coverage.
Note that conditional reweighting does not require data sets (I) and (II) to have identical generators, though they should be as similar as possible to avoid unnecessarily large weights.

\begin{figure}[t]
    \centering
    \includegraphics[width=0.45\textwidth]{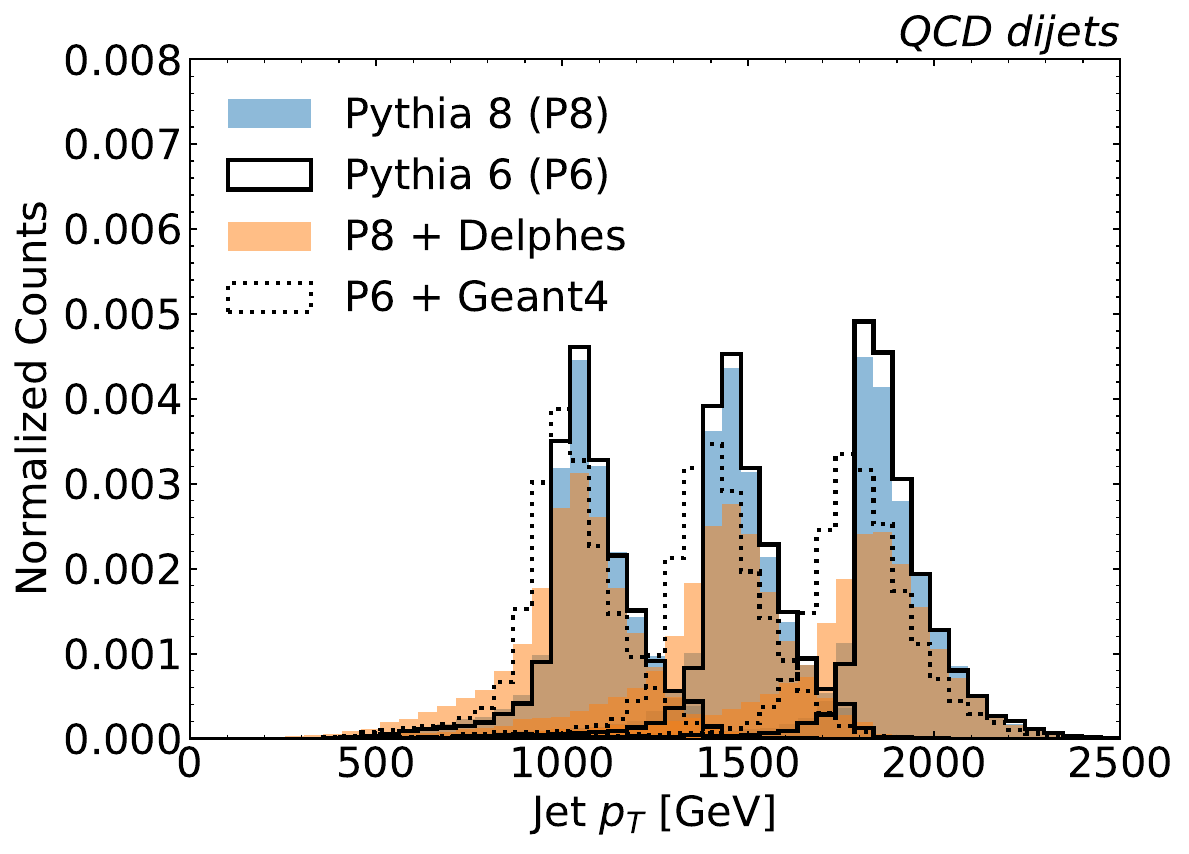}
     \includegraphics[width=0.45\textwidth]{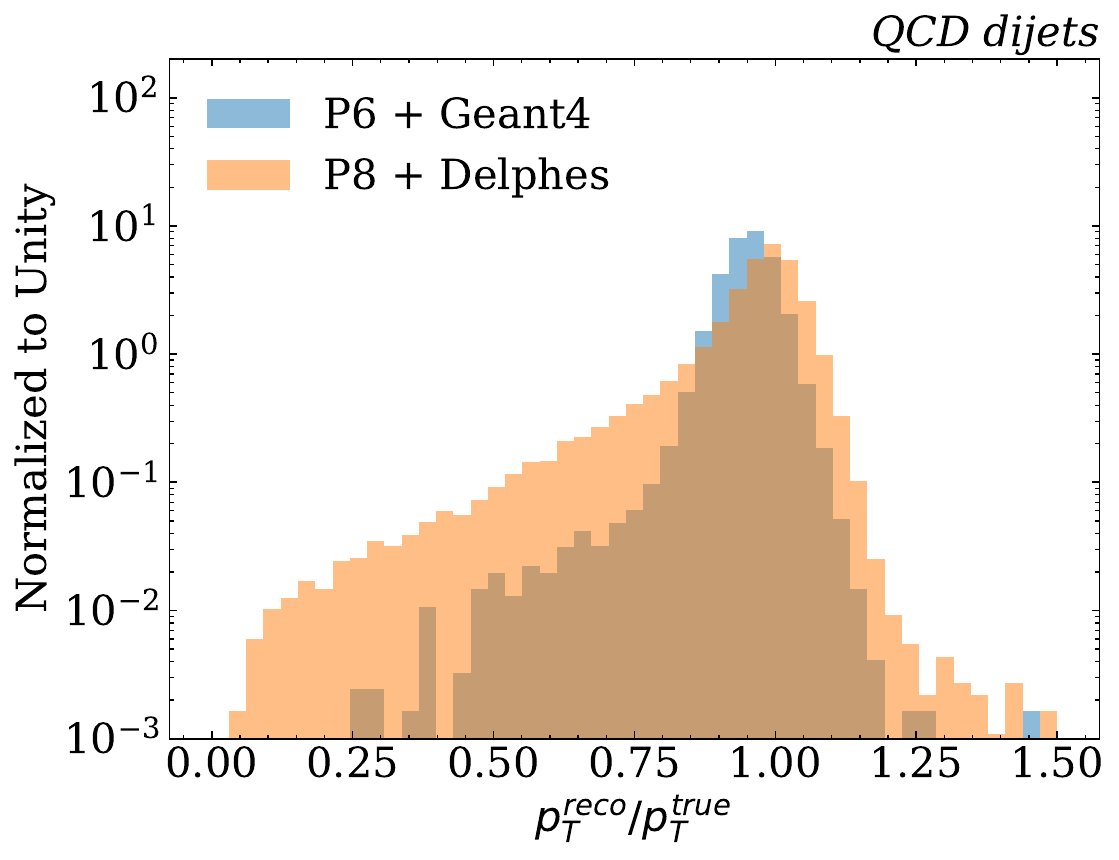}
    \caption{
    Jet kinematics and reconstruction for the QCD example.
    Top: Histograms of particle- and detector-level jet transverse momenta ($p_T$).
    Bottom: Comparing the detector response for the \textsc{Geant4} and \textsc{Delphes} event samples.
    The generation sample for data set (III) is represented by the \textsc{Pythia}~8 (P8) histogram while the corresponding simulation sample is represented by the P8+\textsc{Delphes} histogram.
    The samples for data sets (I) and (II) are missing the middle peak in the top plot.}
    \label{fig:dijet}
\end{figure}

\begin{figure*}[t]
    \centering
    \includegraphics[width=0.4\textwidth]{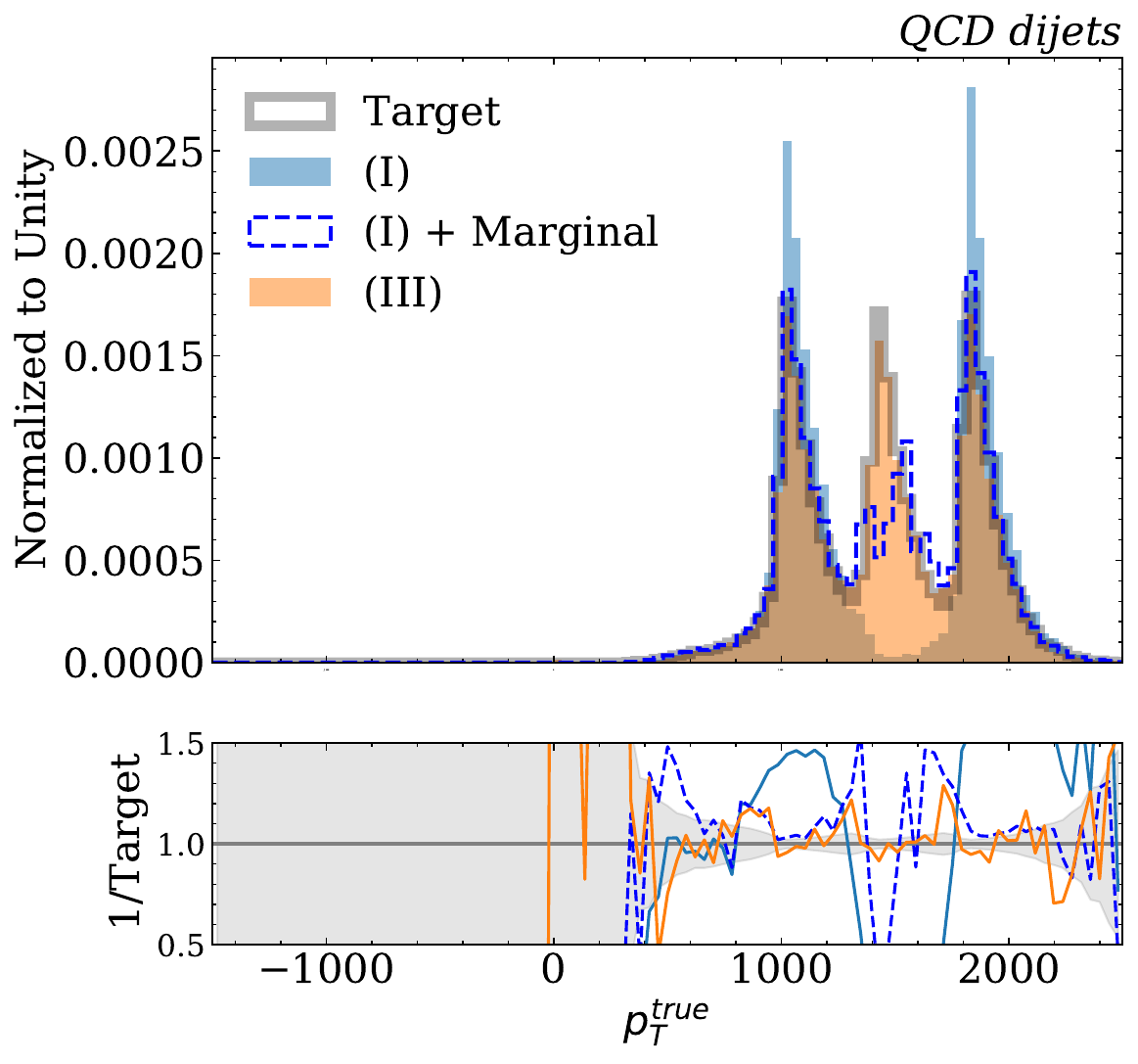}$\qquad$
    \includegraphics[width=0.4\textwidth]{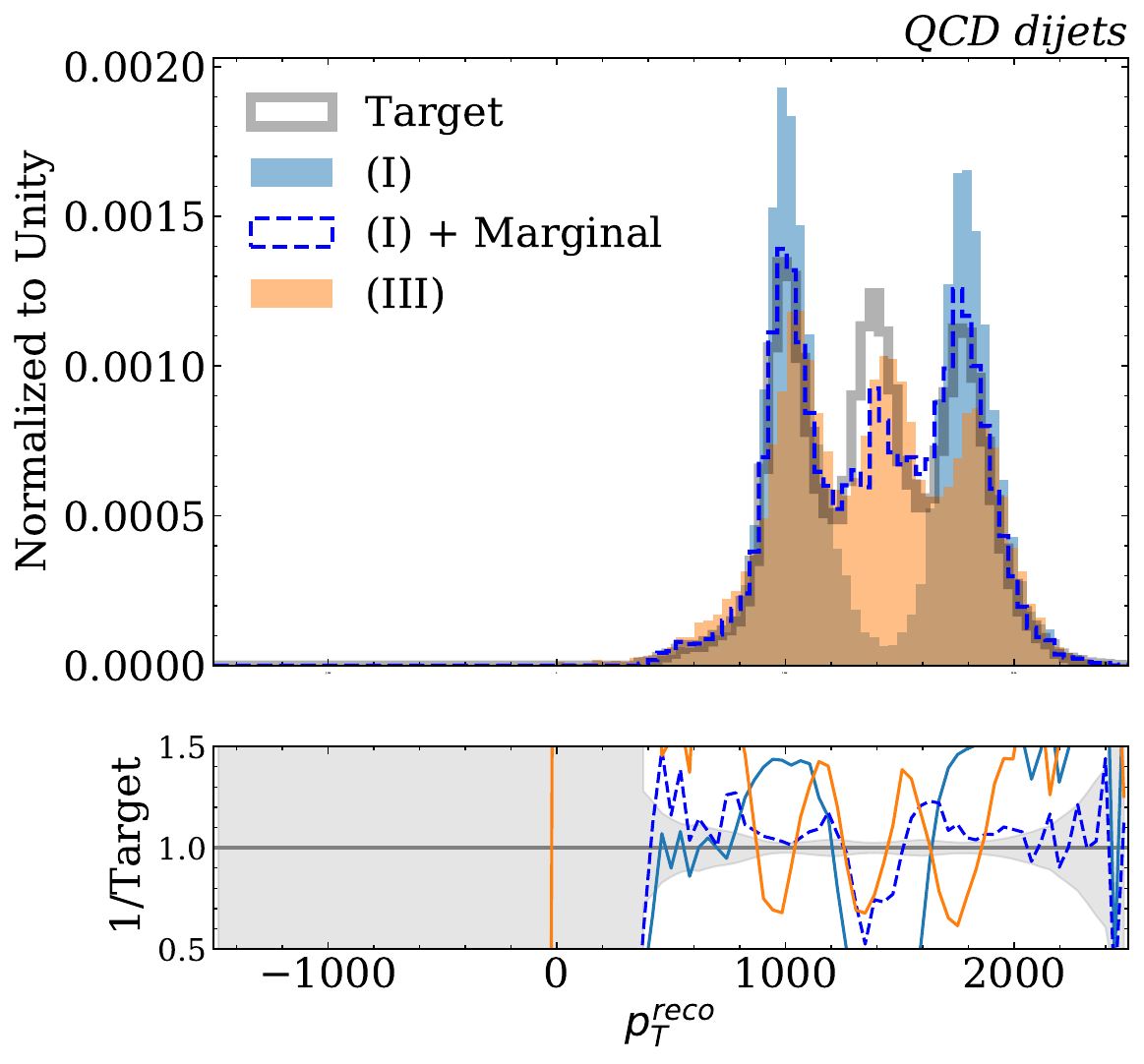}
    
    \includegraphics[width=0.4\textwidth]{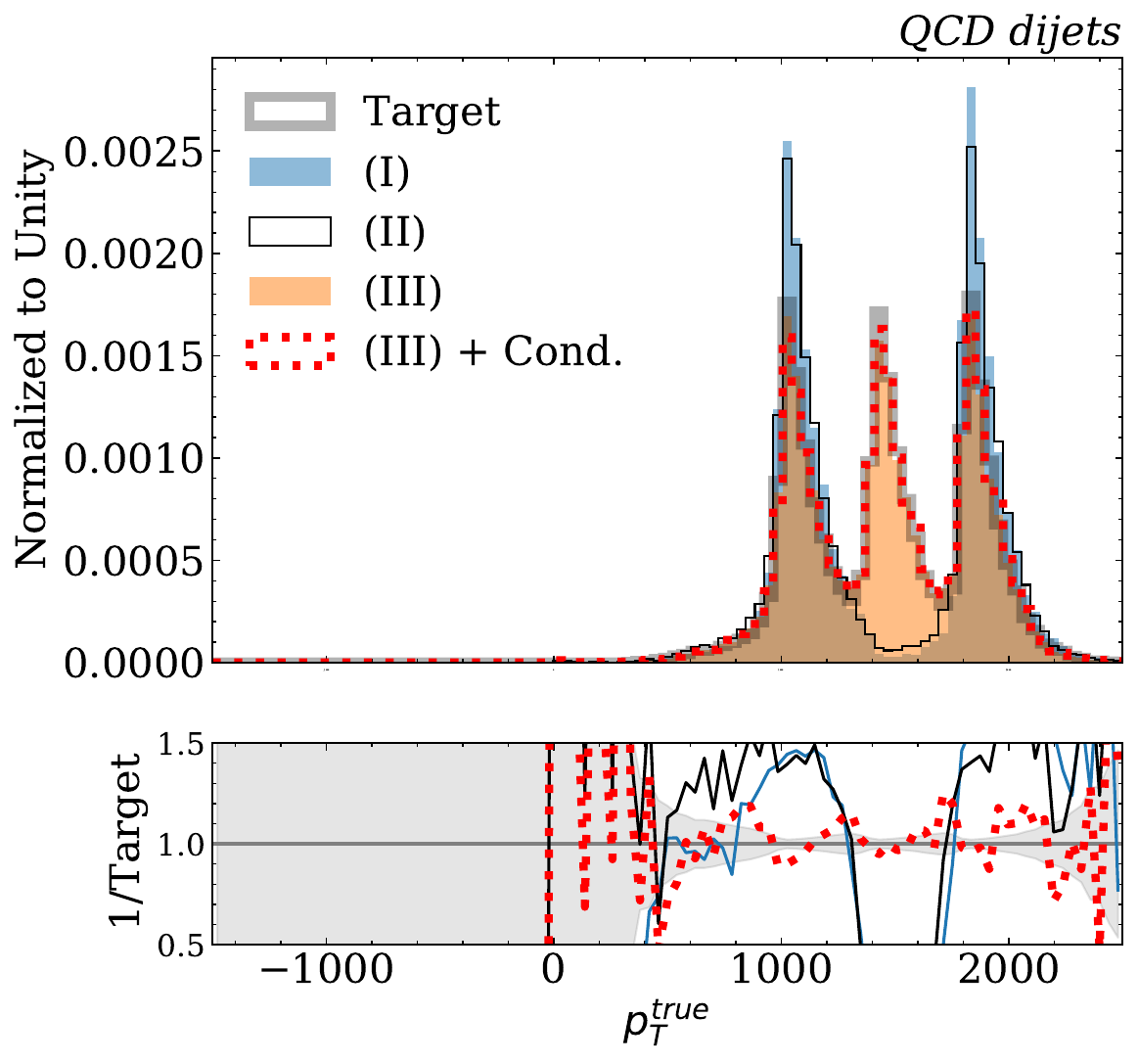}$\qquad$
    \includegraphics[width=0.4\textwidth]{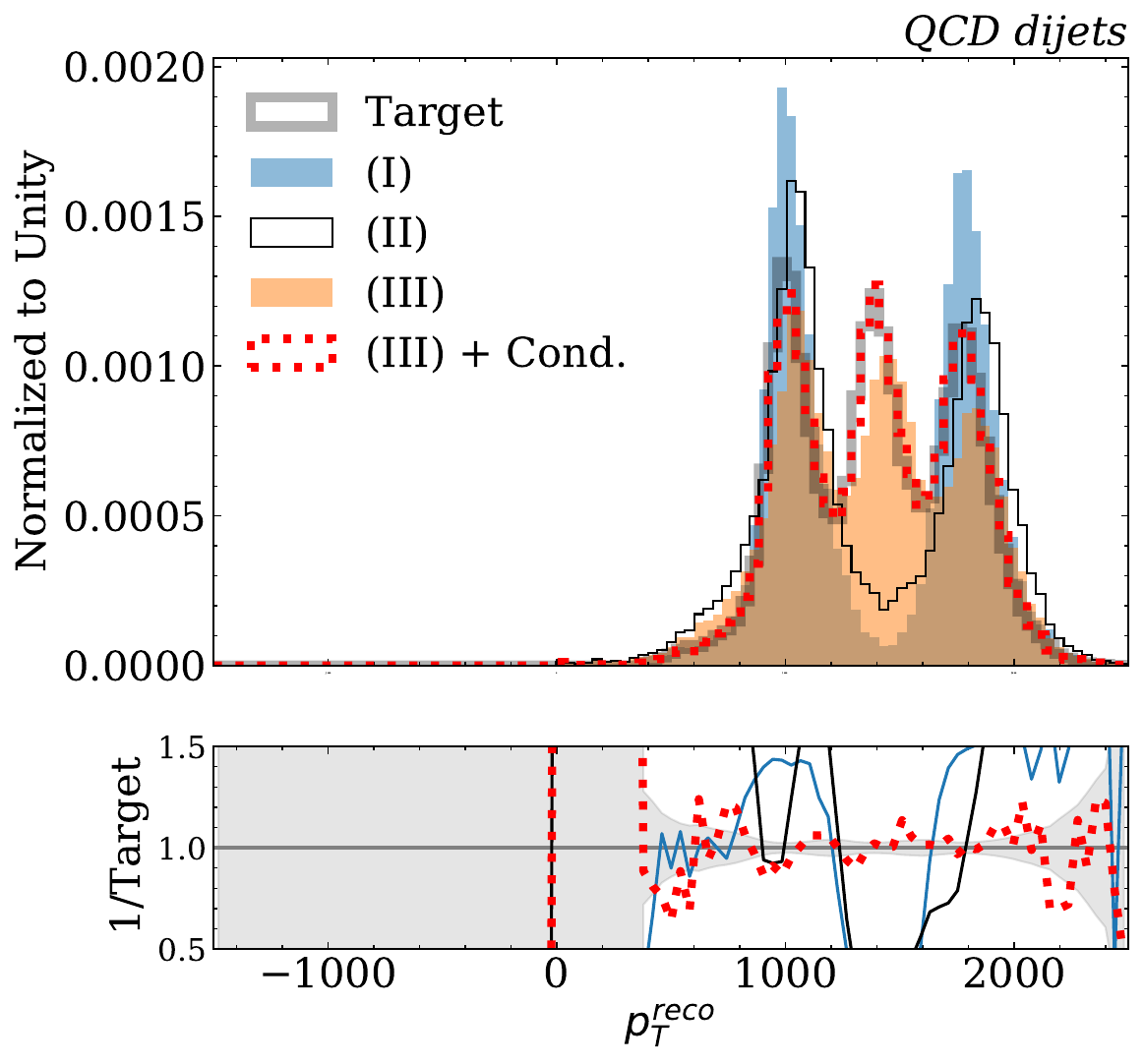}
    \caption{
    Comparison of marginal reweighting (top row) and conditional reweighting (bottom row) in a QCD jet example.
    Shown are histograms of the true particle-level $p_T$ (left column) and reconstructed detector-level $p_T$ (right  column).
    Distribution (I) involves \textsc{Pythia 6} (with an artificial phase space gap at $\hat{p}_T\in[1.4,1.8]~\text{TeV}$) interfaced with \textsc{Geant4}.
    Distribution (II) involves \textsc{Pythia 8} (with the same phase space gap) interfaced with \textsc{Delphes}.
    Distribution (III) involves \textsc{Pythia 8} interfaced with \textsc{Delphes} with no phase space gap.
    To match the target distribution (\textsc{Pythia 8} with no phase space gap interfaced with \textsc{Geant4}), one can either marginally reweight distribution (I) or conditionally reweight distribution (III).
    Like the example in \Fig{gaussianIn}, marginal reweighting cannot bridge the phase space gap, whereas conditional reweighting yields a sensible interpolation.
    }
    \label{fig:dijet2}
\end{figure*}

The fast detector simulation for data sets (II) and (III) is \textsc{Delphes}~3.4.1~\cite{deFavereau:2013fsa,Mertens:2015kba,Selvaggi:2014mya} using the default CMS detector card.
The full detector response for data set (I) uses a \textsc{Geant4}-based~\cite{Agostinelli:2002hh,1610988,Allison:2016lfl} full simulation of the CMS experiment~\cite{Chatrchyan:2008aa}.
More specifically, data set (I) comes from the CMS Open Data Portal~\cite{CMS:QCDsim1000-1400,CMS:QCDsim1400-1800,CMS:QCDsim1800} and processed into an MIT Open Data format~\cite{Komiske:2019jim,komiske_patrick_2019_3341502,komiske_patrick_2019_3341770,komiske_patrick_2019_3341772}.
Data sets (II) and (III) were generated for this study and are available at Ref~\cite{nachman_benjamin_2021_5108967}.

For each data set, we have access to the parton-level hard-scattering scale $\hat{p}_T$ in \textsc{Pythia}, which is in general different from the jet-level transverse momentum $p_T$ we are interested in studying.
As is typical for the generation of steeply falling spectra, the full dijet data sets are constructed as a series of separate samples, each with a different range of $\hat{p}_T$.
To avoid any issues related to the trigger, we focus on data sets where $\hat{p}_T> 1$~TeV.
For this study, we consider three $\hat{p}_T$ ranges:
\begin{equation}
\hat{p}_T\in[1,1.4]~\text{TeV},\quad \hat{p}_T\in[1.4,1.8]~\text{TeV}, \quad \hat{p}_T>1.8~\text{TeV}.
\end{equation}
Particles (at truth level) or particle flow candidates (at reconstructed level) are used as inputs to jet clustering, implemented using \textsc{FastJet}~3.2.1~\cite{Cacciari:2011ma,Cacciari:2005hq} and the anti-$k_t$ algorithm~\cite{Cacciari:2008gp} with radius parameter $R=0.5$.
The corresponding jet $p_T$ spectra are shown in \Fig{dijet}.
When comparing to experimental data, a relative normalization would be applied to scale down the higher $\hat{p}_T$ slices, but we have elided those factors in this study to highlight the behavior of reweighting.

\subsection{Results with Interpolation}

To create a phase space gap and demonstrate the ability of conditional reweighting to interpolate, we remove the $\hat{p}_T\in[1.4,1.8]$~TeV phase space slice from data sets (I) and (II).
This effectively makes them both ``coarse'' generators, relative to the ``precise'' generator for data set (III) that covers the full phase space.
Specifically, our three event samples are
\begin{itemize}
    \item (I): \textsc{Pythia 6} $\Rightarrow$ \textsc{Geant4} for $\hat{p}_T\in[1,1.4]$~TeV and $\hat{p}_T>$~$1.8$~TeV; 
    \item (II): \textsc{Pythia 8} $\Rightarrow$ \textsc{Delphes} for $\hat{p}_T\in[1,1.4]$~TeV and $\hat{p}_T>$~$1.8$~TeV;
    \item (III): \textsc{Pythia 8} $\Rightarrow$ \textsc{Delphes} for $\hat{p}_T >1$~TeV.
\end{itemize}
Each $p_T$ slice within each sample has $10^4$ jets.%
\footnote{The original data sets have many more events, but a relatively small fraction is used here to ensure that the target is more accurate than the reweighted test cases and to amplify the impact of the phase space gap.}
We do not have a simulation of \textsc{Pythia}~8 with \textsc{Geant4}, so unlike the Gaussian case, we cannot display the exact target distribution. 
Instead, we use marginal reweighting on a bigger data set (up to $10^5$ events per $p_T$ slice) without the phase space gap to construct a synthetic target at detector level.

The results of marginal and conditional reweighting for the jet energy response are shown in \Fig{dijet2}.
We use the identical neural network setup from \Sec{gaussian}, and we see the same qualitative features as for the Gaussian interpolation example in \Sec{gauss_interpolation}.
Conditional reweighting correctly has no effect at particle level and yields a smooth distribution at detector level. 
By contrast, marginal reweighting suffers near the phase space gap at 1.5 TeV, similarly to the interpolating Gaussian case.
In addition to better matching the target distribution, conditional reweighting yields weights that are closer to unity.

While we have artificially removed a $\hat{p}_T$ slice for this interpolation study, there are realistic contexts where this could happen.
For example, legacy data corresponding to that $\hat{p}_T$ slice could be missing or corrupted, or that slice may have never been simulated (or simulated with reduced statistics) to save computing power.
%

\section{Conclusions}
\label{sec:conclusions}

In this paper, we extended the technique of neural network-based reweighting to the conditional case, where some features $x$ are reweighted conditioned on other features $x'$.
In regions of phase space that are well covered by the input and target probability densities, conditional reweighting is unlikely to outperform marginal reweighting.
In phase space regions where the input probability density is small compared to the target probability density, though, conditional reweighting can yield improved behavior by leveraging the ability of neural networks to interpolate and extrapolate.
This is relevant for constructing simulated data sets for the LHC, where full simulation may be too computationally costly to cover the full phase space, while fast simulation can be used to fill in the gaps.

An interesting feature of our approach to neural conditional reweighting is that we can derive the reweighting function in \Eq{yprimelimit2} through a single training procedure, instead of the naive two-step procedure suggested by \Eq{conditional_reweighting}.
The key is to train on a higher dimensional phase space and then take an appropriate limit, which may be relevant for other machine learning applications.
In practice, different approaches to neural conditional reweighting yield similar performance, as shown in \App{alt}, but we prefer the single training procedure for its computational simplicity and conceptual elegance.
Though we only showed one-dimensional examples in this paper, there are no conceptual barriers to handling multi-dimensional or variable-dimensional situations, which we plan to explore in future work.

An implicit assumption of our approach is that the neural network is well trained.
This is required for all reweighting methods to work, since the relationship in \Eq{f_to_p} is only guaranteed in the asymptotic limit.
A full quantitative comparison of different neural reweighting methods will need to assess systematic uncertainties, for example by analyzing the results for multiple trainings, performing closure tests on known targets, or comparing the results to low-dimensional binned methods.
Eventually, one might want to use these reweighting uncertainties to guide the process of full simulation, where one prioritizes simulating regions of phase space that cannot be well modeled by (conditional) reweighting alone.

The main advantage of conditional reweighting is in cases where the input probability density is too small relative to the target.
This is a generic challenge, not only for reweighting methods but for any generative modeling task where there is insufficient training data.
Carefully constructed combinations of reweightings may be able to provide a partial solution to this problem, as could imposing smoothness requirements in the loss function to regularize how the neural network interpolates and extrapolates.
Further hybrid methods that involve moving features instead of simply reweighting them (as in optimal transport problems~\cite{Komiske:2019fks,Cai:2020vzx,Romao:2020ojy,Cesarotti:2020hwb,Cesarotti:2020ngq}) may further extend the utility of these methods across high-energy physics and beyond.

\tocless{\section*{Code and Data}}

The code for this paper can be found at \url{https://github.com/hep-lbdl/neuralconditional}, which makes use of \textsc{Jupyter} notebooks~\cite{Kluyver:2016aa} employing \textsc{NumPy}~\cite{harris2020array} for data manipulation and \textsc{Matplotlib}~\cite{Hunter:2007} to produce figures. All of the machine learning was performed on an Nvidia RTX6000 Graphical Processing Unit (GPU) and reproducing the entire notebook takes less than five minutes.  The physics data sets are hosted on Zenodo at~\Ref{komiske_patrick_2019_3341502,komiske_patrick_2019_3341770,komiske_patrick_2019_3341772,nachman_benjamin_2021_5108967}. 

~\\

\begin{acknowledgments}

We thank the anonymous referee for suggesting the use of conditional reweighting in the context of conditional scale factors.
BN is supported by the U.S. Department of Energy (DOE), Office of Science under contract DE-AC02-05CH11231.
JT is supported by the National Science Foundation under Cooperative Agreement PHY-2019786 (The NSF AI Institute for Artificial Intelligence and Fundamental Interactions, \url{http://iaifi.org/}), and by the U.S. DOE Office of High Energy Physics under grant number DE-SC0012567.

\end{acknowledgments}

\FloatBarrier

\begin{figure*}[t]
    \centering     
     \includegraphics[width=0.4\textwidth]{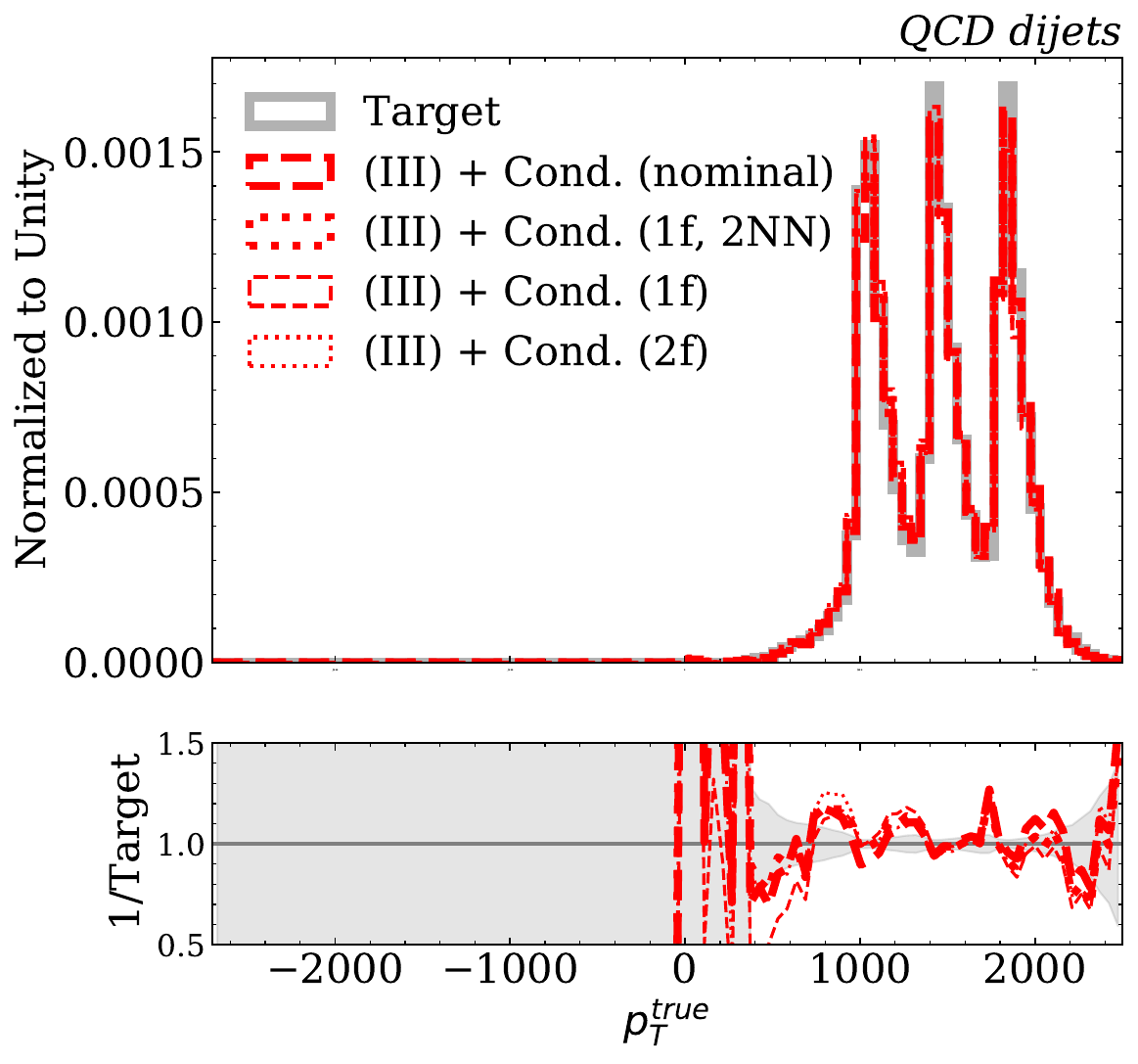}$\qquad$
     \includegraphics[width=0.4\textwidth]{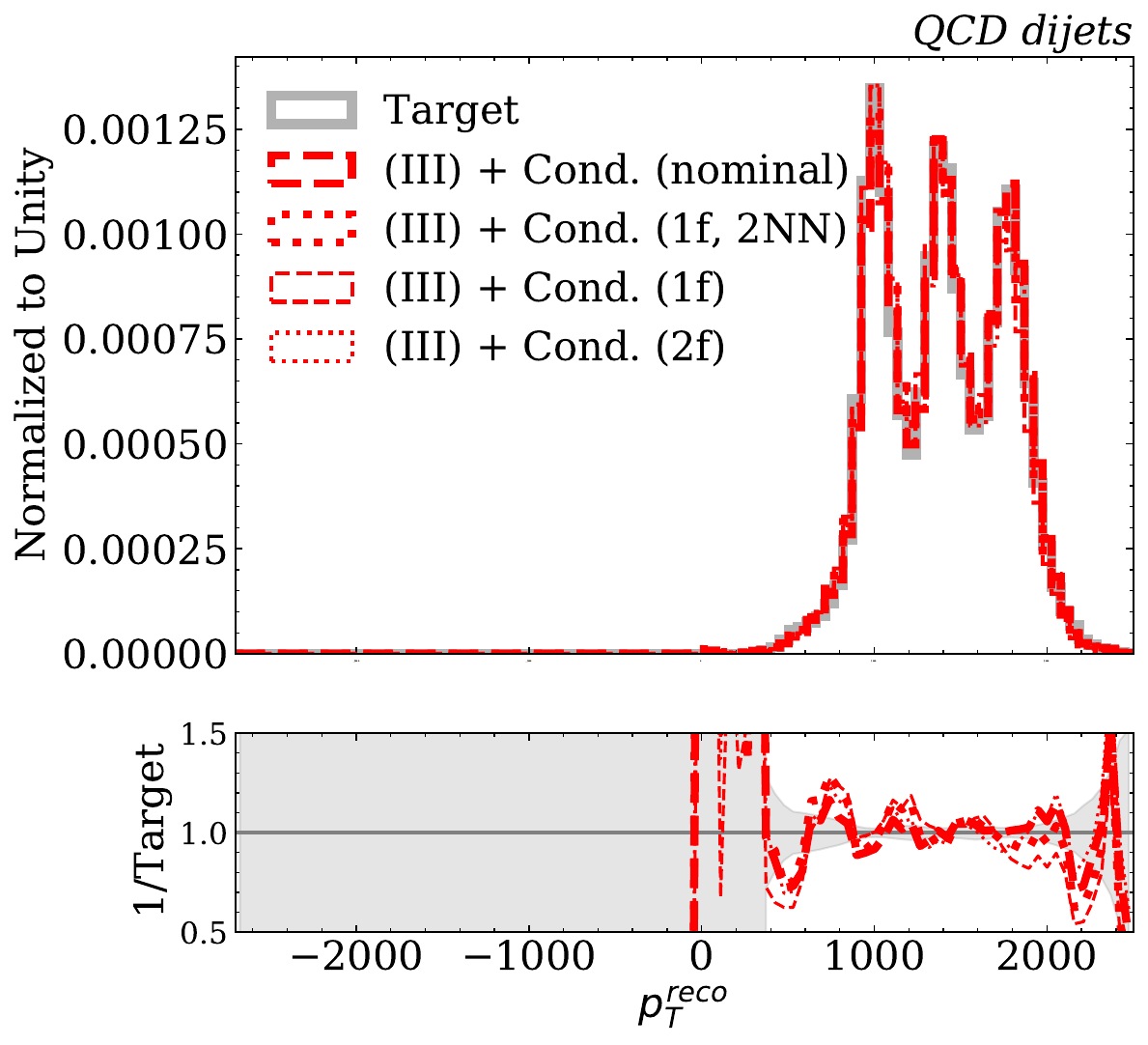}
    \caption{Alternative neural conditional reweighting schemes for the dijet example in \Fig{dijet2}.}
    \label{fig:alt2}
\end{figure*}

\appendix

\section{Alternative Neural Conditional Reweighting Schemes}
\label{app:alt}

In this appendix, we present results for three alternative neural conditional reweighting schemes to explore potential variations.
The methods we compare are:
\begin{itemize}
    \item (nominal):  The nominal scheme, shown in the body of this paper, uses a single learned function built from \Eqss{condtwofuncs}{f0}{f1}.
    \item (1f, 2NN):  This is a slightly more flexible version of the nominal setup, which still uses a single function built from \Eq{condtwofuncs}, but $f_0$ and $f_1$ are now two independent neural networks with the same architecture as the marginal reweighting network.
    \item (1f):  This is an even more flexible setup using the loss in \Eq{ncr}, where we train a single neural network with three inputs without any constraints on its functional form.
    \item (2f):  This is the two function setup from \Eq{conditional_reweighting} that uses one joint reweighting and one marginal reweighting.
\end{itemize}

The results are shown in \Figs{alt2}{alt1}.
All approaches work well, though the nominal approach does a somewhat better job tracking the target distribution.

\begin{figure*}[p]
    \centering
    \includegraphics[width=0.4\textwidth]{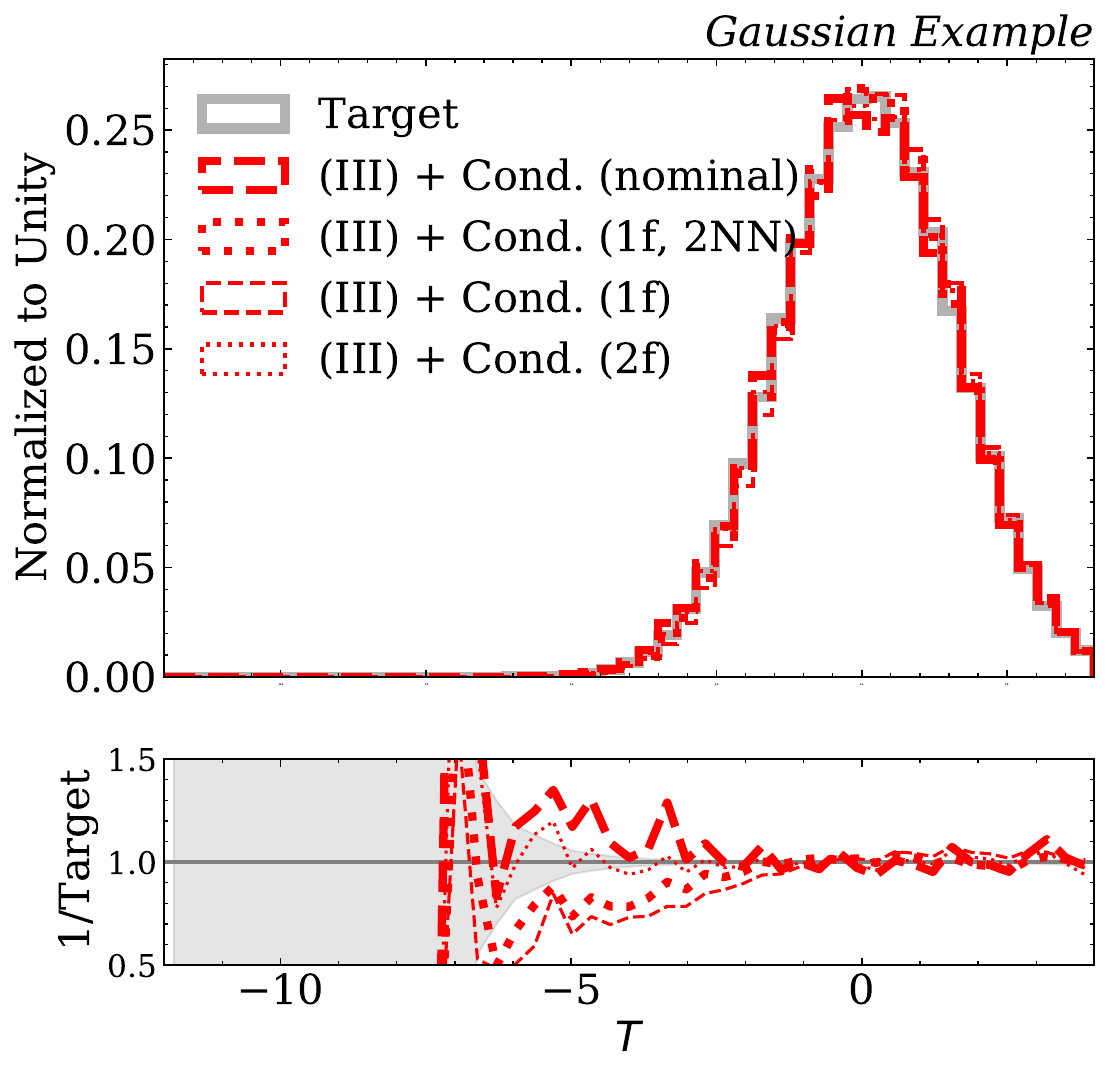} $\qquad$
     \includegraphics[width=0.4\textwidth]{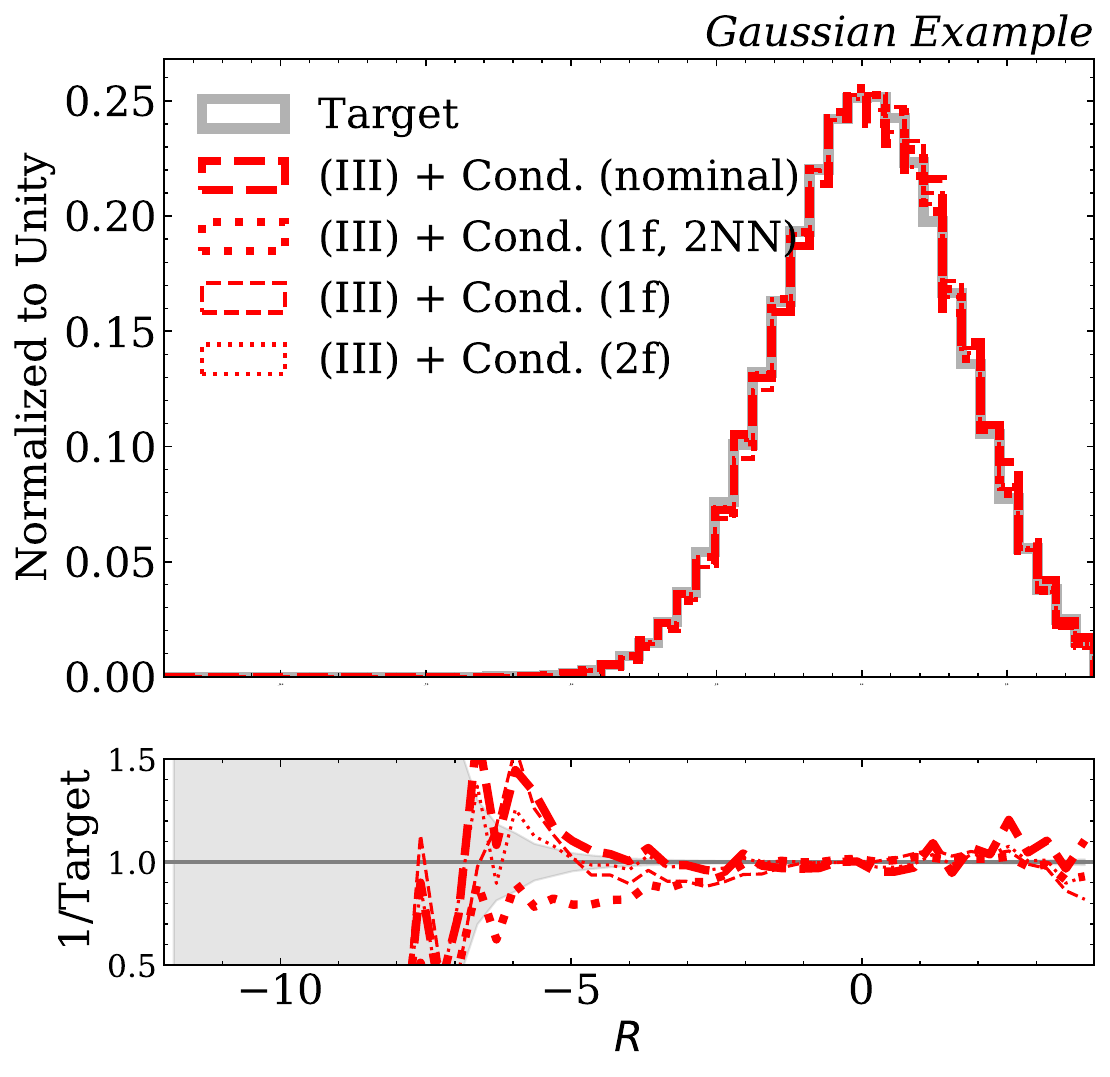}
     
     \includegraphics[width=0.4\textwidth]{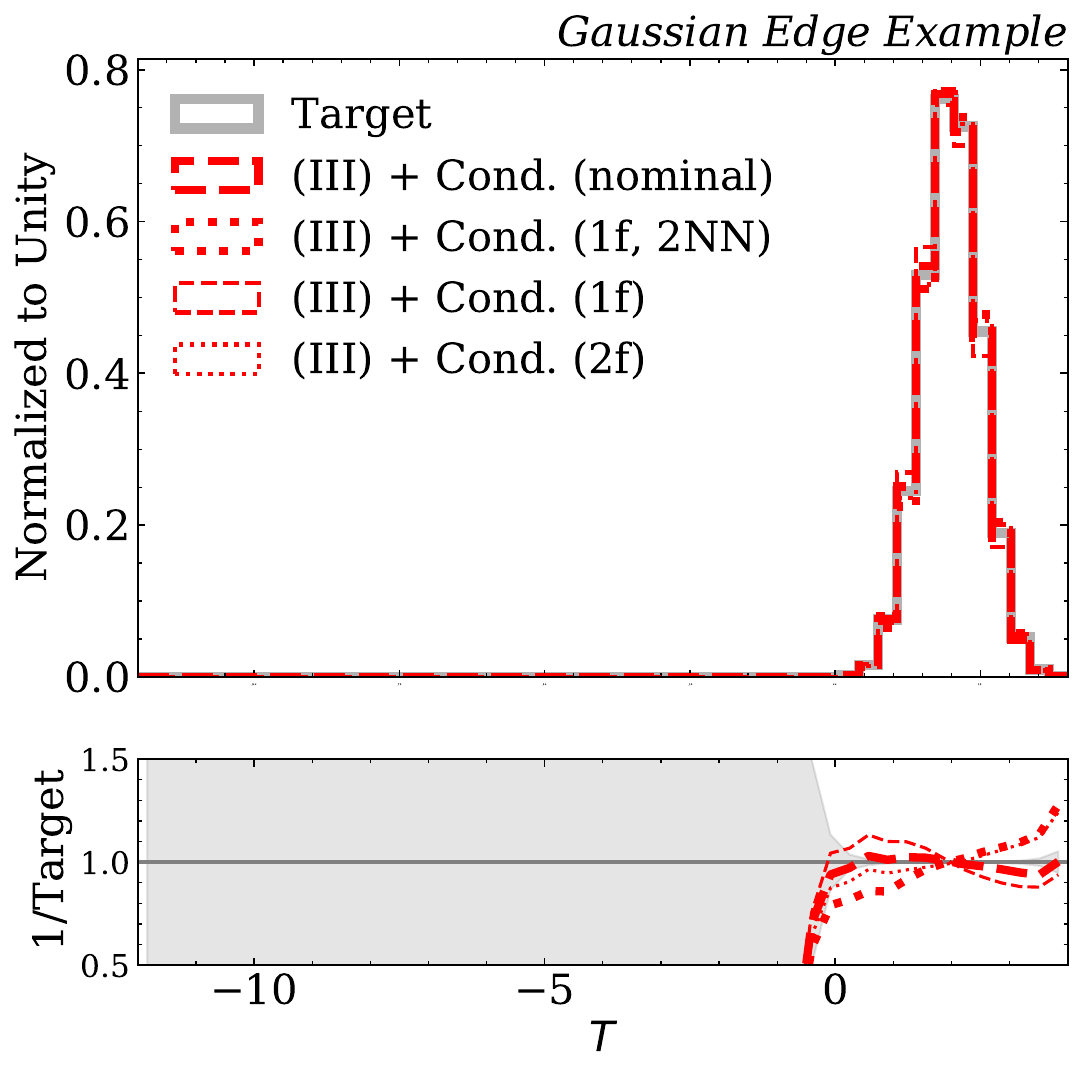}$\qquad$
     \includegraphics[width=0.4\textwidth]{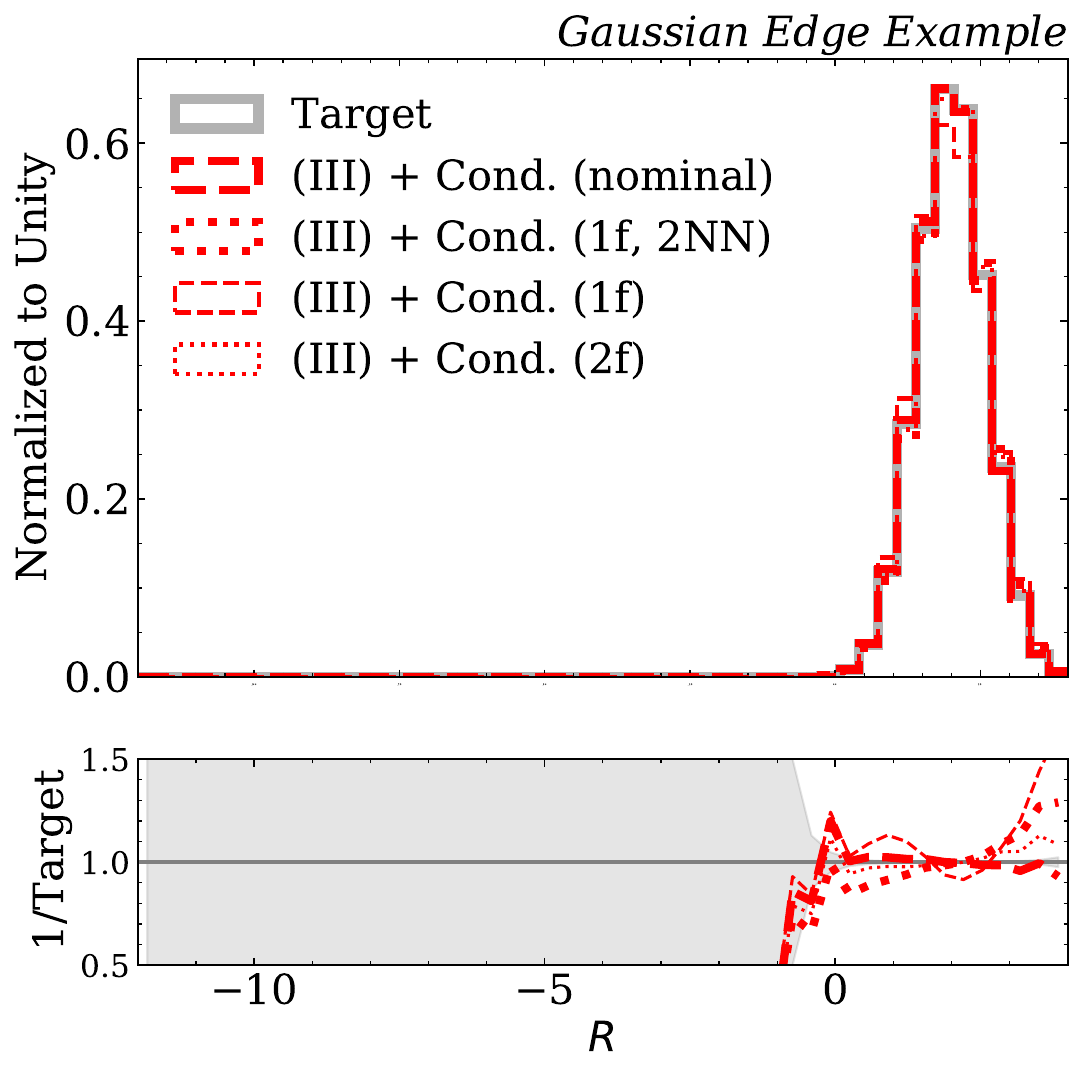}
     
         \includegraphics[width=0.4\textwidth]{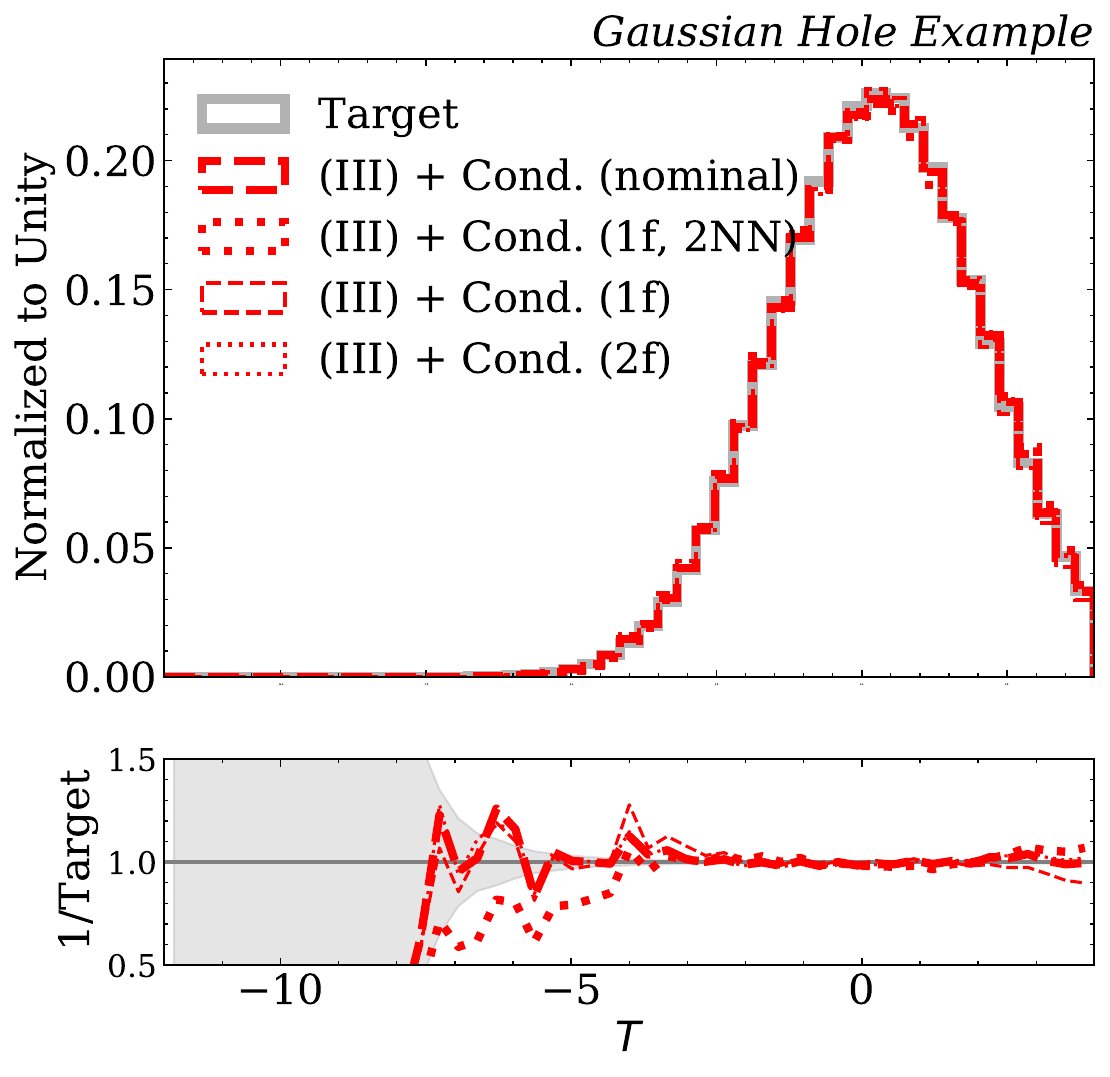}$\qquad$
     \includegraphics[width=0.4\textwidth]{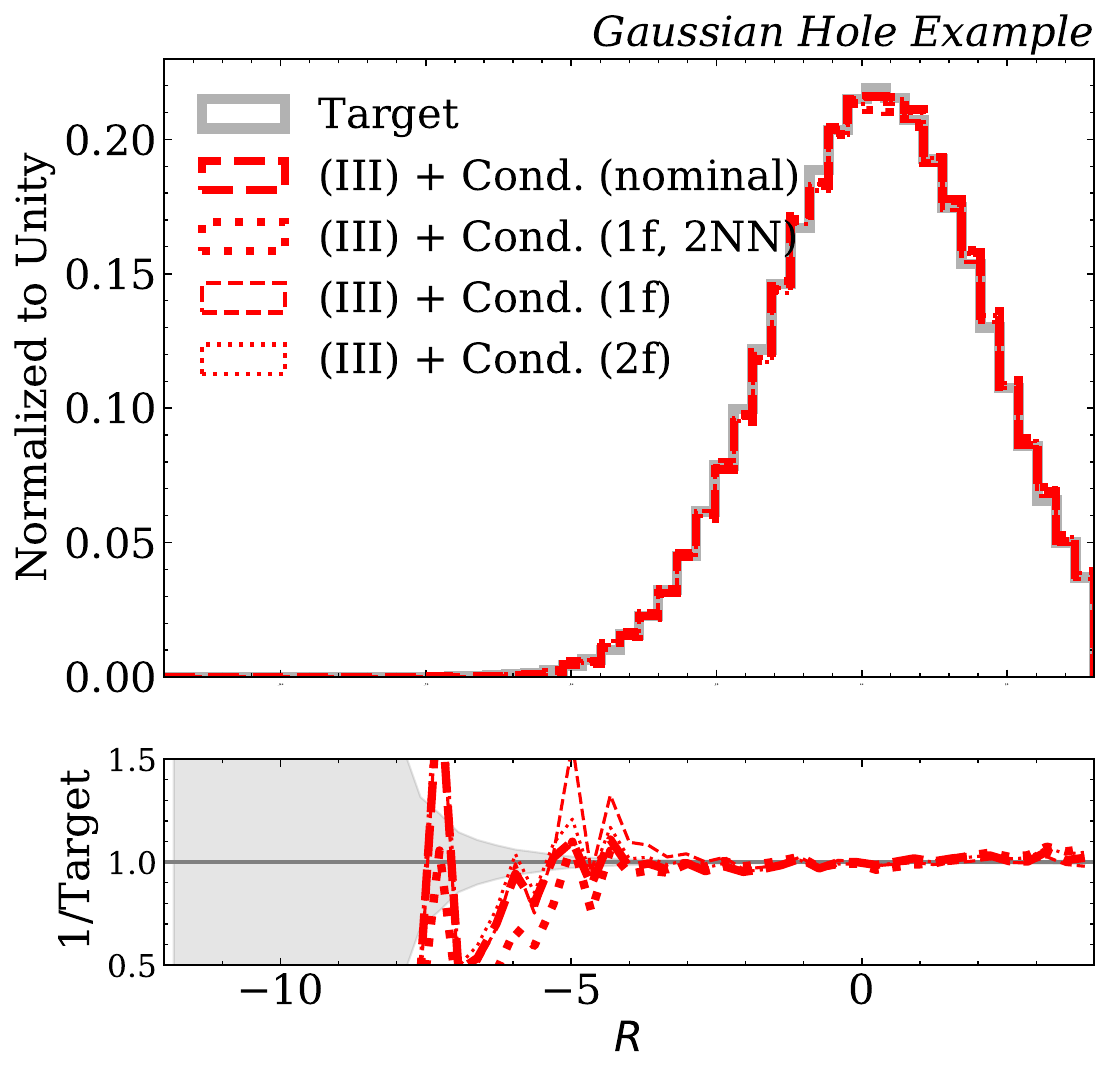}
     
    \caption{Alternative neural conditional reweighting methods for the plain Gaussian example in \Fig{gaussian} (top row), for the extrapolation Gaussian example in \Fig{gaussianEx} (middle row), and for the interpolation Gaussian example in \Fig{gaussianIn} (bottom row).}
    \label{fig:alt1}
\end{figure*}

\bibliography{myrefs}

\end{document}